\documentclass{article}
\usepackage{graphicx} 
\usepackage[a4paper,left=2.73cm,right=2.7cm,top=3cm,bottom=3.5cm]{geometry}

\usepackage{amsfonts,amsmath,amssymb,tabstackengine}
\usepackage[normalem]{ulem}
\usepackage{tikz}
\usepackage{array}
\allowdisplaybreaks
\usepackage{slashed}
\global\long\def\not#1{\slashed{#1}}%
\usepackage{tensor}

\usepackage{comment}

\usepackage{cite}

\definecolor{Gray}{gray}{0.95}

\usepackage[colorlinks=true,linktocpage=true,linkcolor=blue,citecolor=blue]{hyperref}
\usepackage{graphicx}

\usepackage{float}

\addtocontents{toc}{\protect\setcounter{tocdepth}{2}}
\numberwithin{equation}{section}
\usepackage{upgreek}
\usepackage{mathrsfs}
\usepackage{dsfont}
\global\long\def\ri{\mathrm{i}}%
\global\long\def\dd{\mathrm{d}}
\global\long\def\Vol{\mathrm{Vol}}

\global\long\def\AdS{\mathrm{AdS}}%
 \usepackage [latin1]{inputenc}
 
\begin{document}

\begin{titlepage}
	\thispagestyle{empty}
	\begin{center}
	            
	    {\LARGE{\bf  Warped-AdS$_3$ Near-Horizon Geometries\\ \vspace{0,4cm} From TsT Transformations}}
 

 

 
		
		\vspace{50pt}
		
		{S.~Maurelli $^{1,2}$, R.~Noris$^{3}$, M.~Oyarzo$^{4}$, and M.~Trigiante$^{1,2}$}
		
		\vspace{25pt}

		{
		$^1${\it  Department of Applied Science and Technology, Politecnico di Torino, Corso Duca degli Abruzzi, 24, 10129 Torino, Italy}

		\vspace{15pt}

		$^2${\it  INFN, Sezione di Torino, Via P. Giuria 1, 10125 Torino, Italy}

		\vspace{15pt}

		$^3${\it CEICO, Institute of Physics of the Czech Academy of Sciences,
        Na Slovance 2, 182 21 Prague 8, Czech Republic.}

		\vspace{15pt}

		$^4${\it  Departamento de F\'isica, Universidad de Concepci\'on Casilla, 160-C, Concepci\'on, Chile}
		}
		
		\vspace{40pt}
		
		{ABSTRACT}
        
	\end{center}

We provide a string theory realisation of a two-parameter family of Type IIB supergravity solutions of the form ${\rm WAdS}_3 \times {\rm WS}^3 \times M_4$, as near-horizon geometries of supersymmetric backgrounds involving D1-D5 branes, D3-branes and NS5-branes. We show that the full solutions, as well as their near-horizon limits, preserve four supercharges. We end by presenting a new doubly-deformed warped black hole solution and study its thermodynamical properties.
    
\end{titlepage}
\date{January 2025}

\tableofcontents

\section{Introduction}

In the study of AdS/CFT duality, solutions of Type IIB supergravity with spacetime geometry ${\rm AdS}_3\times  S^3\times M_4$, where $M_4$ is either $T^4$ or $K_3$, have played a special role for several reasons. Such backgrounds can be supported purely by NSNS fields, and a worldsheet description of superstring theory on them in terms of a ${\rm SL}(2,\mathbb{R})\times {\rm SU}(2)$-WZW model is available and has been extensively studied \cite{Deger:1998nm,deBoer:1998kjm,Maldacena:2000hw,Maldacena:2000kv,Maldacena:2001km}. This makes it possible, in certain limits, to test the holographic duality beyond the supergravity approximation \cite{Eberhardt:2017fsi,Eberhardt:2019ywk}. 

Moreover, this geometry locally describes the near-horizon limit of the so-called D1-D5 black holes, for which the first microscopic entropy counting within superstring theory was achieved \cite{Strominger:1996sh,Strominger:1997eq,Maldacena:1998bw}. The 1+1 SCFT dual to Type IIB superstring theory on ${\rm AdS}_3\times  S^3\times M_4$ is a deformation of a sigma model on ${\rm Sym}^N(M_4)$, whose symmetry is the asymptotic symmetry group of the ${\rm AdS}_3$ factor  \cite{Brown:1986nw}, namely the infinite-dimensional (Virasoro)$\times$(Virasoro) extension of the ${\rm SL}(2,\mathbb{R})_L \times {\rm SL}(2,\mathbb{R})_R$ isometry group of the same space.
\par 
Special deformations of these backgrounds are obtained by ``warping'' the ${\rm AdS}_3$ factor and/or the $\mathrm{S}^3$ one, the resulting spaces being denoted by ${\rm WAdS}_3$ and ${\rm WS}^3$, respectively. The deformation of the former space leads to a dual field theory which is no longer conformal and has been identified with a so-called \emph{dipole}-theory, arising from a suitable \emph{irrelevant} deformation of the D1-D5 SCFT. Despite the reduced amount of symmetry, certain backgrounds of the form ${\rm WAdS}_3\times  {\rm WS}^3\times M_4$ (with the warping on one of the first two factors or on both) have been shown to describe integrable marginal deformations of the superstring theory on the undeformed background \cite{Israel:2004vv,Detournay:2005fz,Hoare:2022asa}. The key to establishing this property is the fact that some of these backgrounds can be obtained through a TsT duality. The latter is an ${\rm O}(d,d)$ transformation, generalizing the ${\rm O}(d,d;\mathbb{Z})$ symmetry of the worldsheet theory on ${\rm AdS}_3\times S^3\times M_4$, that preserves the integrability of the model \cite{Orlando:2019rjg}. Other backgrounds of the form ${\rm WAdS}_3\times  {\rm WS}^3\times M_4$, some of which are supersymmetric, were recently constructed in \cite{Maurelli:2025iba}, without resorting to TsT transformations. The novel feature of these solutions is their dependence on two independent parameters, which control the warping of the two factors separately. 

The existence of such solutions in Type IIB supergravity naturally raises the question regarding their string-theoretic origin: can these geometries be described as an appropriate near-horizon limit of some brane configuration, as in the standard D1-D5 brane system? In this paper, we address this question by explicitly constructing doubly-deformed Type IIB supergravity solutions, involving D1-D5 branes, standard D3 branes and NS5 branes. These configurations interpolate between an asymptotically locally flat spacetime at radial infinity and null-$\mathrm{WAdS}_3\times \mathrm{WS}^3\times T^4$ spacetime in the near-horizon region. Despite their complexity, we find that the interpolating backgrounds, as well as their near-horizon limits, are $1/8$ BPS, thus preserving four supercharges. Interestingly, unlike in the standard D1-D5 case, these deformed solutions do not exhibit a symmetry enhancement in the near-horizon limit. They are generated via double-TsT transformations from the D1-D5 solution; their supersymmetry contradicts the expectation that double-TsT would break all supersymmetries of the $\mathrm{AdS}_3\times \mathrm{S}^3\times T^4$ configuration \cite{Hoare:2022asa}, while this possibility was simply not considered in \cite{Orlando:2010ay}, where the authors focused on spacelike $\mathrm{WAdS}_3$. Related constructions in Type IIB supergravity, based on TsT transformations and dualities, were explored, for the case $M_4 = K_3$, in \cite{Georgescu:2024iam}.
Remarkably, the interpolating solutions constructed here, all describe Kundt spacetimes with a null Killing vector (to be referred to as \emph{Killing-Kundt}), though not pp-wave geometries. This is related to the corresponding TsT transformation being made along a null direction of ${\rm AdS}_3$.

These new understandings suggest that all ${\rm WAdS}_3\times  {\rm WS}^3\times M_4$ solutions should arise as the near-horizon limits of suitable brane configurations. However, constructing the explicit brane setup is not always straightforward, particularly when the warping is spacelike or timelike. The string theoretic origin of such configurations becomes even more relevant in the presence of spacelike warpings, where the black hole configuration can be obtained, in certain cases, by compactifying along appropriate directions. For this reason, we construct a novel doubly-deformed, warped-AdS-black hole (warped black hole in short) solution, whose brane interpretation is left for a future investigation.

The paper is organised as follows: in section \ref{sec2}, we review the main properties of the standard D1-D5 system in the case $M_4=T^4$, setting the stage for the results presented in later sections. In section \ref{sec3}, we present the two doubly-deformed solutions describing different brane configurations in Type IIB supergravity, study their geometrical properties, their supersymmetry, and finally consider an appropriate near-horizon limit yielding a ${\rm WAdS}_3\times  {\rm WS}^3\times T^4$ geometry. In section \ref{sec4}, we generate a warped black hole configuration with two independent parameters and study its first law of thermodynamics. We end with some concluding remarks in section \ref{conclusions}. In Appendix \ref{appIIB}, we review the conventions used for Type IIB supergravity, whereas Appendix \ref{appTsT} contains, for the reader's convenience, a short review of the TsT solution generating technique. Finally, Appendices \ref{proofsusy} and \ref{proofsusy2} are devoted to the details about the derivation of the Killing spinors for two of the solutions we discuss.

\section{Review of the D1-D5 brane system}\label{sec2}

Let us recall the well-known D1-D5 configuration of Type IIB supergravity, whose main definitions and equations are listed in Appendix \ref{appIIB}. The stack of overlapping D5 branes shares one spatial direction with the D1 branes, while the other are wrapped on $T^4$. Such a solution can be expressed, in the string frame and in terms of $[r, x_+, x_-, \psi, \varphi_1 , \varphi_2, y^1, y^2, y^3, y^4]$ coordinates, as
\begin{align}\label{D1D5sol}
\dd s^{2} & =\sqrt{H_{1}H_{5}}\left(\dd r^{2}+\frac{r^{2}}{4}[\dd\psi^{2}+\sin^{2}\psi\dd\varphi_{1}^{2}+(\dd\varphi_{2}-\cos\psi\dd\varphi_{1})^{2}]\right)-\frac{\dd x_{+}\dd x_{-}}{\sqrt{H_{1}H_{5}}}+\sqrt{\frac{H_{1}}{H_{5}}}\dd\overset{\to}{y}{}^2\,,\nonumber\\
\Phi & =\frac{1}{2}\log\frac{H_{1}}{H_{5}}\,,\nonumber\\
\mathsf{F}_{3} & =\frac{1}{8}d_{5}\sin\psi\dd\varphi_{1}\wedge\dd\varphi_{2}\wedge\dd\psi+\frac{d_{1}}{2r^{3}H_{1}^{2}}\dd r\wedge\dd x_{-}\wedge\dd x_{+}\,,\nonumber\\
\mathsf{H}_{3} & =0\,,\quad\mathsf{F}_{1}=0\,,\quad\mathsf{F}_{5}=0\,, \nonumber\\
H_{i} & =1+\frac{d_{i}}{2r^{2}}\,,
\end{align}
where the ranges of the coordinates are:
$$r\in \mathbb{R}_+,\,x_\pm\in \mathbb{R}\,\,,\,\,\,\psi\in [0,\,\pi],\,\varphi_1\in [0,\,2\pi],\,\,\varphi_2\in [0,\,4\pi)\,,\,\,y_{k}\in [0,\,2\pi)\,,$$
$y_k$ ($k=1,2,3,4$) being the coordinates of $T^4$. \par
The Ramond-Ramond charges associated with $\mathsf{F}_{3}$ and its dual $\mathsf{F}_{7}=-\star\mathsf{F}_{3}$ are given by
\begin{align}
\frac{1}{2\pi^{2}}\int_{S^{3}}\mathsf{F}_{3} & =d_{5}\,,\nonumber\\
\frac{1}{(2\pi)^{4}\,2\pi{}^{2}}\int_{S^{3}\times T^{4}}\mathsf{F}_{7} & =d_{1}\,,
\end{align}
and are proportional to the number of D5 and D1 branes, respectively, while the brane configuration is illustrated in table \ref{fig:D1D5}.
\begin{figure}[H]
    \centering

\begin{center}
\begin{tabular}{c|c|c|c|c|c|c|c|c|c|c}
 & $x_{+}$ & $x_{-}$ & $r$ & $\psi$ & $\varphi_1$ & $\varphi_2$ & $y_{1}$ & $y_{2}$ & $y_{3}$ & $y_{4}$\tabularnewline
\hline 
D1 & $\times$ & $\times$ &  &  &  &  &  &  &  & \tabularnewline
\hline 
D5 & $\times$ & $\times$ &  &  &  &  & $\times$ & $\times$ & $\times$ & $\times$\tabularnewline
\end{tabular}
\par\end{center}
    
    \caption{In the above table, $\times$ indicates the directions along the D-branes, while empty spaces are the transverse directions. }
    \label{fig:D1D5}
\end{figure}

The solution preserves eight supercharges. Indeed, considering the following vielbein basis
\begin{align}
    e^0&=\frac{\dd x_++\dd x_-}{2(H_1 H_5)^\frac14}\,,\qquad e^1=\frac{\dd x_+-\dd x_-}{2(H_1 H_5)^\frac14}\,,\qquad e^2=(H_1 H_5)^\frac14\dd r\,,\nonumber\\
    e^3&=\frac{r}{2}(H_1 H_5)^\frac14\dd\psi\,,\qquad e^4=\frac{r}{2}(H_1 H_5)^\frac14\dd\varphi_1 \sin\psi\,,\qquad e^5=\frac{r}{2}(H_1 H_5)^\frac14(\dd\varphi_2-\cos\psi\dd\varphi_1)\,,\nonumber\\
    e^6&=\left(\frac{H_1}{H_5}\right)^\frac14\dd y_1\,, \qquad e^7=\left(\frac{H_1}{H_5}\right)^\frac14\dd y_2\,, \qquad e^8=\left(\frac{H_1}{H_5}\right)^\frac14\dd y_3\,, \qquad e^9=\left(\frac{H_1}{H_5}\right)^\frac14\dd y_4\,, \qquad
\end{align}
the Killing spinor is given by 
\begin{align}
    \epsilon & =\frac{1}{(4H_1H_5)^\frac18}e^{-\frac{\psi}{4}(\Gamma^{45}-\Gamma^{23})}e^{-\frac{\varphi_{1}}{4}(\Gamma^{25}-\Gamma^{34})}e^{\frac{\varphi_{2}}{4}(\Gamma^{25}+\Gamma^{34})}P_{1}P_{2}P_{0}\epsilon_{0}\,,\nonumber\\
    P_2 &= \frac12(\mathsf{1}_{32}\otimes \mathsf 1_2+\Gamma^{2345}\otimes \sigma_1)\,,\nonumber\\
    P_1 &= \frac12(\mathsf{1}_{32}-\Gamma^{012345})\otimes \mathsf 1_2\,,\nonumber\\
    P_0 &= \frac12(\mathsf{1}_{32}+\Gamma_*)\otimes \mathsf 1_2\,,
\end{align}
with $\epsilon_0$ a constant spinor. 

Finally, the near-horizon geometry of such a configuration is obtained by rescaling the following coordinates simultaneously
\begin{align}\label{nearhorD15}
    r\to \lambda r\,, \qquad x_+\to \frac{\sqrt{d_1d_5}}{\lambda} x_+\,, \qquad x_-\to \frac{\sqrt{d_1d_5}}{\lambda} x_-
\end{align}
and by subsequently considering the $\lambda\to0$ limit. Let us emphasise the isotropy of such rescaling, which acts on both $x_+$ and $x_-$ in the same way. The resulting solution reads
\begin{align}
    \dd s^2&=\frac{\sqrt{d_1d_5}}{2}\left(\frac{\dd r^2}{r^2}-4r^2 \dd x_+\dd x_-+\frac14[\dd\psi^{2}+\sin^{2}\psi\dd\varphi_{1}^{2}+(\dd\varphi_{2}-\cos\psi\dd\varphi_{1})^{2}]\right)+\sqrt\frac{d_1}{d_5}\dd\overset{\to}{y}{}^2\,,\nonumber\\
    \Phi & =\frac12 \log\frac{d_1}{d_5}\,,\nonumber\\
    \mathsf{F}_{3} & =d_5\left(2 r \dd r \wedge \dd x_-\wedge\dd x_++\frac18\sin\psi\dd\psi\wedge \dd \varphi_1\wedge \dd \varphi_2\right)\,,\nonumber\\
    \mathsf{H}_{3} & =0\,,\quad\mathsf{F}_{1}=0\,,\quad\mathsf{F}_{5}=0\,.
\end{align}
The metric describes an $\AdS_3\times \mathrm{S}^3\times T^4$ geometry, with enhanced isometries and supersymmetries with respect to the original D-brane system.

\section{Lightlike double deformations of the D1-D5 brane system}\label{sec3}

In this section, we will generate Type IIB supergravity backgrounds through the TsT duality generating solution technique \cite{Lunin:2005jy, Catal-Ozer:2005dux, Castellani:2024ial}, see  Appendix \ref{appTsT} for a concise review. In the following, $(x_1,x_2,\gamma)$ will denote a TsT duality with T along the coordinate direction $x_1$ and shift s along $x_2$, with parameter $\gamma$. Explicitly, this is given by the action, in  the worldsheet theory, of the ${\rm O}(d,d)$ element resulting from the following consecutive transformations:
\begin{align}
    x_1\to \tilde x_1\,\qquad x_2\to x_2-\gamma \tilde x_1\,,\qquad \tilde x_1 \to \hat x_1\,.
\end{align} 
In the following subsections, we will apply two  distinct double TsT dualities on the D1-D5 brane configuration, thus introducing two free parameters, and study the resulting geometries. 

\subsection{Supersymmetric Killing-Kundt deformation}

The solution obtained through a double TsT along  $(x_+, y_4, \gamma_1), \, (\varphi_2, y_3, \gamma_2)$ reads
\begin{align}\label{firstTsT}
\dd s^{2} & =\sqrt{H_{1}H_{5}}\left[\dd r^{2}+\frac{r^{2}}{4}\left(\dd\psi^{2}+\sin^{2}\psi\,\dd\varphi_{1}^{2}+\frac{H_{5}}{H_{1}}e^{2\Phi}\mathbf{A}_{2}^{2}\right)\right]-\sqrt{\frac{H_{5}}{H_{1}}}\left(\mathbf{A}_{1}\dd x_{+}+\frac{\gamma_{1}^{2}}{4}\mathbf{A}_{1}^{2}\right)\nonumber \\
& +\sqrt{\frac{H_{1}}{H_{5}}}\left(\dd y_{1}^{2}+\dd y_{2}^{2}+\frac{H_{5}}{H_{1}}e^{2\Phi}\dd y_{3}^{2}+\dd y_{4}^{2}\right)\,, \nonumber\\
\mathsf B_{2} & =\frac{\gamma_{1}}{2}\mathbf{A}_{1}\wedge\dd y_{4}-\frac{H_{5}\gamma_{2}r^{2}}{4}e^{2\Phi}\mathbf{A}_{2}\wedge\dd y_{3}\,,\qquad e^{2\Phi}=\frac{4H_{1}}{(4+r^{2}\gamma_{2}^{2}H_{1})H_{5}}\,, \nonumber\\
\mathsf{F}_{3} & =\frac{1}{8}(d_{5}\sin\psi\dd\varphi_{1}\wedge\dd\varphi_{2}\wedge\dd\psi+\frac{4d_{1}}{r^{3}H_{1}^{2}}\dd r\wedge\dd x_{-}\wedge\dd x_{+})\,,\qquad\mathsf{F}_{1}=0\,, \nonumber\\
\mathsf{F}_{5} & =\frac{\gamma_{1}d_{5}}{2}\mathbf{A}_{1}\wedge\left(\frac{1}{8}\sin\psi\mathbf{A}_{2}\wedge\dd y_{4}\wedge\dd\varphi_{1}\wedge\dd\psi+\frac{1}{r^{3}H_{5}}\dd r\wedge\dd y_{1}\wedge\dd y_{2}\wedge\dd y_{3}\right) \nonumber\\
& +\frac{\gamma_{2}d_{1}}{8}\left(-\sin\psi\dd y_{1}\wedge\dd y_{2}\wedge\dd y_{4}\wedge\dd\varphi_{1}\wedge\dd\psi+\frac{H_{5}^{2}}{rH_{1}^{2}}e^{2\Phi}\mathbf{A}_{1}\wedge\mathbf{A}_{2}\wedge\dd r\wedge\dd x_{+}\wedge\dd y_{3}\right)\,,
\end{align}
where we defined the 1-forms
\begin{align}\label{eq:A1=000020null=000020case1}
\mathbf{A}_{1} & =\frac{\dd x_{-}}{H_{5}}\,,\nonumber\\
\mathbf{A}_{2} & =\dd\varphi_{2}-\cos\psi\dd\varphi_{1}\,.
\end{align}
They satisfy a Beltrami-like equation (see e.g. \cite{Andrianopoli:2023dfm}) with non-constant mass, when restricted to corresponding 3-dimensional subspaces. Indeed, we find
\begin{align}
\star^{(r,x_\pm)}\dd\mathbf{A}_{1} & =\frac{2(1-H_{5})}{rH_{1}^{1/4}H_{5}^{5/4}}\mathbf{A}_{1}\,,\qquad \star^{(\psi,\varphi_{1,2})}\dd\mathbf{A}_{2} =\frac{4}{r (H_1H_5)^{\frac14}\sqrt{4+r^2\gamma_2^2H_1}}\mathbf{A}_{2}\,,
\end{align}
where $\star^{(r,x_\pm)}$ and $\star^{(\psi,\varphi_{1,2})}$ are the induced Hodge stars in the respective 3-dimensional
spaces.  

The solution depends explicitly on $\mathbf{A}_{2}$, which is ill-defined at the poles of $S^2$ located at $\psi=0,\pi$. However, one can shift $\varphi_2\to \varphi_2\pm \varphi_1$, yielding 
\begin{align}
     \mathbf{A}_{2} \to \mathbf{A}_{2}^\pm=\dd\varphi_2-(\cos\psi\mp 1)\dd\varphi_1\,,
\end{align}
where the upper and lower signs regularise the solution in the north and south poles, respectively. The term in $\mathsf H_3$ depending on $\mathbf{A}_{2}$ signals the presence of a source for an NS5 brane which is smeared over $r$. 

\begin{figure}[H]
    \centering
    \includegraphics[width=0.8\linewidth]{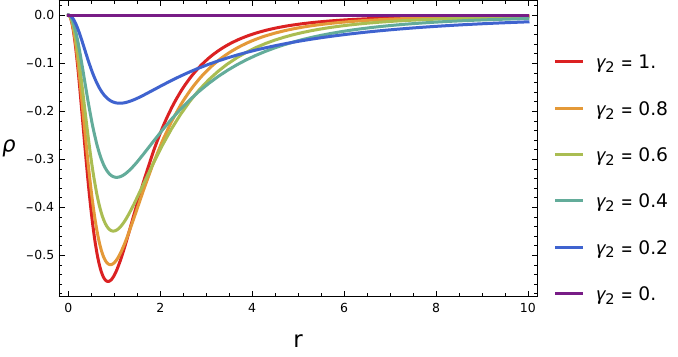}
    \caption{The charge density $\rho(r,\psi)$ for different values of $\gamma_2$, given $d_1=1.5$, $d_5=1$, $\psi=\frac{\pi}{2}$.}
    \label{Figure1}
\end{figure}

The total solution \eqref{firstTsT} describes a D1-D5 system in the presence of NS5-branes and standard D3 branes. In particular, the space transverse to the D1-D5 branes system corresponds to a four-dimensional space with a radial coordinate $r$. Differently from the D1-D5 system, such a space is not flat, but it contains a squashed 3-sphere whose deforming parameter depends on the radial coordinate. \\
The full 10-dimensional spacetime has no curvature singularities, as it is well behaved in $r=0$. Indeed, the Riemann tensor in the vielbein basis is everywhere finite in such a limit. In the $r\to\infty$ limit, the space is asymptotically locally flat, as the curvature in rigid components vanishes and the dilaton field behaves as \footnote{The limits $r\to\infty$ and $\gamma_2\to0$ do not commute. Taking the latter limit first, leads to a divergent dilaton, whereas the former leads to a vanishing one.}
\begin{align}
g_{s}=e^{\Phi} & =\frac{2}{\gamma_{2}r}+O(r^{-3})\,.
\end{align}
The charges associated with the brane configurations are as follows
\begin{align}\label{charges first tst}
    \frac{1}{2\pi^{2}}\int_{S^{3}}\mathsf{F}_{3} & =d_{5}\,,\nonumber\\
    \frac{1}{(2\pi)^{4}\,2\pi{}^{2}}\int_{S^{3}\times T^{4}}(\mathsf{F}_{7} -\mathsf B_2\wedge \dd C_4)& =d_{1}\,, \nonumber\\
    \frac{1}{(2\pi)^3 4\pi }\int_{(S^1)^3\times S^2}(\mathsf F_5-\mathsf B_2\wedge \mathsf F_3) &= \frac{d_1 \gamma_2}{8}\,.
\end{align}
As the NS5-brane is smeared over the radial coordinate, the associated charge will depend on $r$. To see this, let us consider the terms in $\mathsf{H}_3$ that have a compact support
\begin{align}
    \mathsf{H}_3 = f(r)\sin\psi \dd\psi\wedge \dd \varphi_1\wedge \dd y_3+\ldots\equiv \hat{\mathsf{H}}_3+\ldots\,,
\end{align}
where we defined $f(r)=-\frac{\gamma_{2}}{4}e^{2\Phi(r)}r^{2}H_{5}$ and $\ldots$ indicate forms with non-compact support. Now let us consider the integral of the above expression on a domain $\Sigma_r=S^2(\psi,\varphi_1)\times S^1(y_3)$ at radius $r$. Let us define the charge as 
\begin{align}\label{chargeNS5}
    \frac{q^{\mathrm{NS5}}(r)}{8\pi^2}&=\int_{\Sigma_r}\hat{\mathsf{H}}_3-\int_{\Sigma_0}\hat{\mathsf{H}}_3= \int_{D} \dd \hat{\mathsf{H}}_3 \nonumber\\
    &= \int_{D}  f'(r)\sin\psi \dd r\wedge \dd\psi\wedge \dd \varphi_1\wedge \dd y_3 = \int_D \rho(r,\psi) \Vol(D)\,,
\end{align}
where $D =\Sigma(\psi,\varphi_1,y_3) \ltimes [0,r]$ and $\Vol(D)$ is the induced volume four-form on $D$. The explicit expression of the NS5 charge density is given by 
\begin{align}
    \rho(r,\psi)=\frac{e^{2\Phi}H_5^\frac12}{rH_1^2}\frac{4\gamma_2\sin\psi}{\sqrt{4+r^2\sin^2\psi\gamma_2^2H_1}}\,,
\end{align}
and it is plotted, for different values of the deforming parameter $\gamma_2$, in Figure \ref{Figure1}. The average value of $r$ defined by this density is given, for $\gamma_2\neq0$, by 
\begin{align}
    \langle r\rangle=\pi  \sqrt{\frac{d_1}{8} +\frac{1}{\gamma^2_2}}\,.
\end{align} 
From \eqref{chargeNS5}, one can see that the charge is finite for any value of $r$, including the limiting value of $r\to \infty$, while the above Figure shows that the corresponding density is localised. The brane configuration is illustrated in table \ref{fig:firstTsT}. The solution contains the brane configuration illustrated in table \ref{fig:firstTsT}.

\begin{figure}[H]
    \centering

\begin{center}
\begin{tabular}{c|c|c|c|c|c|c|c|c|c|c}
 & $x_{+}$ & $x_{-}$ & $r$ & $\psi$ & $\varphi_1$ & $\varphi_2$ & $y_{1}$ & $y_{2}$ & $y_{3}$ & $y_{4}$\tabularnewline
\hline 
D1 & $\times$ & $\times$ &  &  &  &  &  &  &  & \tabularnewline
\hline 
D5 & $\times$ & $\times$ &  &  &  &  & $\times$ & $\times$ & $\times$ & $\times$\tabularnewline
\hline 
D3 & $\times$ & $\times$ &  &  &  & $\times$ &  &  & $\times$  \tabularnewline
\hline 
NS5 & $\times$ & $\times$ &  &  &  & $\times$ & $\times$  & $\times$ &  & $\times$ \tabularnewline
\end{tabular}
\par\end{center}
    
    \caption{In the above table, the $\times$ along the $\varphi_2$ direction actually indicates that the corresponding extended objects is wrapped the $\mathrm{S}^1$ dual to $\mathbf{A}_2$. The longitudinal directions for each extended object are identified as the dual to the ones defining the corresponding charges.}
    \label{fig:firstTsT}
\end{figure}

Finally, let us observe that the solution discussed above can also be described within the $D=6,\,\mathcal{N}=(1,1)$ supergravity considered in \cite{Maurelli:2025iba}, by compactifing on $T_4/\mathbb Z_2$. Interestingly, notice that, on the deformed D1-D5 background, some of the ${\rm SO}(4,4)/{{\rm SO}(4)\times {\rm SO}(4)}$ scalars of the model evolve, just as for the solution of \cite{Georgescu:2024iam}, toward the origin of the scalar manifold in the IR.
 
\paragraph{Killing-Kundt class}

The geometry appearing in \eqref{firstTsT} describes a Kundt spacetime. Indeed, let us consider the following vectors
\begin{align}\label{nullvectorsKundt}
    \ell^{\mu}\partial_{\mu} & =-2\partial_+\,,\nonumber\\
    k^{\mu}\partial_{\mu} & =\frac{1}{4} \gamma^2_1 \sqrt{\frac{H_1}{H_5}}\partial_+-\sqrt{H_1H_5}\partial_-\,, \nonumber\\
    \ell^{\mu}k_{\mu} & =-1\,,\qquad\ell^{\mu}\ell_{\mu}=0\,,\qquad k^{\mu}k_{\mu}=0\,,
\end{align}
with $\ell^\mu$ geodesic, that is
\begin{align}\label{geodesic}
\qquad\ell^{\mu}\nabla_{\mu}\ell^{\nu}=0\,.
\end{align}
The above vector fields allow us to define the following rank-8 projector
\begin{align}
P^\mu{}_{\nu} & =\delta^\mu_{\nu}+k^{\mu}\ell_{\nu}+\ell^{\mu}k_{\nu}\,.
\end{align}
The twist, expansion and shear are defined as the irreducible components
of the tensor field \cite{Griffiths:2009dfa}
\begin{align}
T_{\mu\nu}^{\perp\perp} & =P_{\mu}{}^{\rho}P_{\nu}{}^{\sigma}\nabla_{\rho}\ell_{\sigma}\,.
\end{align}
For the solution under consideration, this tensor vanishes identically, which implies that the spacetime is Kundt. Moreover, the null and geodesic vector field $\ell$ is also Killing, leading to a further restriction of the Kundt class of spacetimes, which we refer to as Killing-Kundt. Observe that this does not imply that $\ell$ is covariantly constant, since the change of $\ell$ in the direction $k$, measured by $k^\mu\nabla_\mu \ell^\nu$, does not vanish.

\paragraph{Supersymmetry}
The solution \eqref{firstTsT} preserves four out of the initial eight supercharges. Indeed, given the vielbein basis:
\begin{align}
    e^{0} & =\dd x_{+}+\frac{\dd x_-}{4H_5}\left(\gamma_1^2+\sqrt{\frac{H_5}{H_1}}\right)\,,\qquad e^{1}=\dd r (H_1 H_5)^{\frac14}\,,\qquad e^{2}=\dd x_{+}+\frac{\dd x_-}{4H_5}\left(\gamma_1^2-\sqrt{\frac{H_5}{H_1}}\right)\,,\nonumber \\
    e^{3} & =\frac r2\dd\psi(H_1 H_5)^{\frac14}\,,\qquad e^{4}=\frac r2\sin\psi\dd\varphi_{1}(H_1 H_5)^{\frac14}\,,\qquad e^{5}=\frac{r}{2}e^\Phi \frac{H_5^\frac34}{H_1^\frac14} (\dd\varphi_{2}-\cos\psi\dd\varphi_{1})\,,\\
    e^{6} & =\left(\frac{H_1}{H_5}\right)^\frac14\dd y^{1}\,,\qquad e^{7}=\left(\frac{H_1}{H_5}\right)^\frac14\dd y^{2}\,,\qquad e^{8}=\left(\frac{H_5}{H_1}\right)^\frac14e^\Phi\dd y^{3}\,,\qquad e^{9}=\left(\frac{H_1}{H_5}\right)^\frac14\dd y^{4}\,,\nonumber
\end{align}
the Killing spinor reads
\begin{align}\label{KSfirsttst}
    \epsilon=\frac12\left(\frac{d_{1}+2r^{2}}{8+(d_{1}+2r^{2})\gamma_{2}^{2}}\right)^{1/4}e^{\frac{\psi}{2}\Gamma^{13}}e^{-\frac{\varphi_1}{2}\Gamma^{15}\sigma_1}\left(\alpha_{+}-\alpha_{-}\mathbf{\Gamma}_{\mathtt{2}}\right)(1-\mathbf{\Gamma}_{\mathtt{1}})P_{1}P_{2}P_0\epsilon_0\,,
\end{align}
where
\begin{align}
\boldsymbol{\Gamma}_{\mathtt{1}} & =\sigma_{1}\,,\qquad\mathbf{\Gamma}_{\mathtt{2}}=\Gamma^{58}\ri\sigma_{2}\,,\qquad \mathsf{x}(r) =-\frac{2\sqrt{2}}{\sqrt{8+d_{1}\gamma_{2}^{2}+2r^{2}\gamma_{2}^{2}}}\,, \qquad \mathsf{y}(r)  =\frac{\sqrt{2r^2+d_{1}}\gamma_{2}}{\sqrt{8+d_{1}\gamma_{2}^{2}+2r^{2}\gamma_{2}^{2}}}\,,\nonumber\\
\alpha_{\pm} & =\sqrt{\frac{1\pm\mathsf{x}(r)}{\mathsf{y}(r)}}\,, \qquad P_1=\frac{1}{2}(1-\Gamma^{02})\,,\,\qquad P_2=\frac{1}{2}(1+\sigma_{1}\Gamma^{6789})\,.
\end{align}
We refer the reader to Appendix \ref{proofsusy} for a  detailed of derivation of eq. \eqref{KSfirsttst}.

\paragraph{Near-horizon geometry analysis}
Let us finally analyze the near-horizon geometry of this solution. To this end, let us rescale the coordinates as
\begin{align}\label{nearhorizonlimit}
r & =\lambda\tilde{r}\,,\qquad x_{-}=\xi\tilde{x}_{-}\,.
\end{align}
We then fix $\xi$ in terms of $\lambda$ so that, in the $\lambda\to0$ limit, $\mathbf{A}_{1}$ does not vanish. Explicitly, we have
\begin{align}
\mathbf{A}_{1} & =\frac{\xi\dd\tilde{x}_{-}}{1+\frac{d_{5}}{2\lambda^{2}\tilde{r}^{2}}}\to\frac{2\xi\lambda^{2}}{d_{5}}\tilde{r}^{2}\dd\tilde{x}_{-}\,,
\end{align}
which requires $\xi\lambda^{2}=d_{5}k$. Let us remark here that, differently from the near-horizon limit considered in \eqref{nearhorD15} for the D1-D5 system, the above rescaling is ``non-relativistic'', as it only involves $x_-$, while $x_+$ is unaffected. Such chiral rescaling is deeply related to the nature of the resulting geometry, which, as we will see, features a $\mathrm{WAdS}_3$ factor. The latter is indeed a chiral deformation of $\AdS_3$, which breaks one of the two $\mathrm{SL}(2,\mathbb R)$ factors in the $\AdS_3$ isometry to $\mathrm U(1)$.

If we relabel $\tilde r=r$, $\tilde x_-=x_-$, rescale $\gamma_2\to 8 \gamma_2$, choose $k=d_1 d_5$ and define the near horizon 1-forms as
\begin{align}\label{1formnearhorizon}
\mathsf{A}_{1} & ={r}^{2}\dd {x}_{-}\,\nonumber\\
\mathsf{A}_{2} & =\mathbf{A}_{2}\,,
\end{align}
the solution reduces to 
\begin{align}\label{firstTsTnearhorizon}
\dd s^{2} & =\frac{\sqrt{d_{1}d_{5}}}{2}\left(\frac{\dd {r}^{2}}{r^2}-4r^2 \dd x_-\dd x_{+}+\frac{1}{4}[\dd\psi^{2}+\sin^{2}\psi\dd\varphi_{1}^{2}+(\dd\varphi_{2}-\cos\psi\dd\varphi_{1})^{2}]\right)\,\nonumber\\
&-\sqrt{d_{1}d_5}\left(d_1\gamma^{2}_1\mathsf{A}_{1}^2+d_5\gamma_2^2e^{2\Phi}\mathsf{A}_2^2\right)+\sqrt{\frac{d_{1}}{d_{5}}}\left(\dd y_1^2+\dd y_2^2+\frac{d_5}{d_1}e^{2\Phi}\dd y_3^2+\dd y_4^2\right)\,,\nonumber\\
\mathsf B_{2} & =\gamma_1 d_1\mathsf{A}_{1}\wedge\dd y^{4}-d_5\gamma_2e^{2\Phi}\mathsf{A}_2\wedge \dd y^3\,,\qquad\Phi=\frac{1}{2}\log\frac{d_{1}}{d_{5}(1+8d_1\gamma_2^2)}\,,\nonumber\\
\mathsf{F}_{3} & =\frac{1}{8}d_{5}\sin\psi\dd\varphi_{1}\wedge\dd\varphi_{2}\wedge\dd\psi+2 d_5 r\dd r\wedge\dd x_{-}\wedge\dd x_{+}\,,\qquad\mathsf{F}_{1}=0\,,\nonumber\\
\mathsf{F}_{5} & =\gamma_{1}d_1\mathsf{A}_{1}\wedge\left(\frac{d_{5}}{8}\sin\psi\mathsf{A}_2\wedge\dd y^{4}\wedge\dd\varphi_{1}\wedge\dd\psi+\frac{2}{{r}}\dd r\wedge\dd y^{1}\wedge\dd y^{2}\wedge\dd y^{3}\right)\,\nonumber\\
&+d_5\gamma_2e^{2\Phi} \left(-\frac{2  d_5 \mathsf{A}_1\wedge \dd r\wedge \dd x_+\wedge \dd y^3\wedge }{r}+\frac{d_1}{d_5}e^{-2\Phi}\dd y^1\wedge \dd y^2\wedge \dd y^4\wedge \dd \right)\mathsf{A}_2\,.
\end{align}
The total metric describes a lightlike-$\mathrm{WAdS}_3\times \mathrm{WS}^3\times T^4$ spacetime, where both $\AdS_3$ and $\mathrm{S^3}$ factors of the near-horizon geometry of the D1-D5 system have been warped\footnote{We refer the reader to Section 3 of \cite{Maurelli:2025iba} for a concise, unifying review of the WAdS$_3$ and WS$_3$ spaces and for the definition of the notations used here for their description. In particular their characterization in terms of the 1-forms ${\bf A}_1,\,{\bf A}_2$}. The $\mathrm{WAdS}_3$ and $\mathrm{WS}^3$ radii coincide and are given by $L=\left(\frac{d_1d_5}{4}e^{-\Phi}\right)^\frac14$. Interestingly, this explicit value depends on one of the warping parameters, $\gamma_2$. 

The $1$-forms in \eqref{1formnearhorizon} satisfy Beltrami-like equations with constant mass parameters, the sign of the latter determining the chirality of the deformation, that is which of the left/right $\mathrm{SL}(2,\mathbb R)$ and $\mathrm{SU}(2)$ factors of the undeformed configuration are broken to $\mathrm{U}(1)$. The total isometry group is therefore given by $(\mathrm{SL}(2,\mathbb R)\times \mathrm{U}(1))\times (\mathrm{SU}(2)\times \mathrm{U}(1))\times \mathrm{U}(1)^4$. This finally shows that the warped $\AdS$ geometry arises as a non-relativistic near-horizon limit of the locally asymptotically flat solution given in \eqref{firstTsT}. The latter therefore interpolates between $\mathrm{WAdS}_3\times \mathrm{WS}^3\times T^4$ and an asymptotically locally flat spacetime, which is realised in perturbative string theory. This provides us with a valuable handle on the microscopic description of this background and of its dual theory.

Finally, let us comment on the supersymmetry of the obtained solution: interestingly, \eqref{firstTsTnearhorizon} does not feature a supersymmetry enhancement, contrary to the undeformed D1-D5 case, as it preserves four supercharges, as the starting configuration \eqref{firstTsT}.
Given the following vielbein, 
\begin{align}
        e^{0} & =\dd x_{+}+\frac{\sqrt{d_1}r^2}{2}\dd x_-(d_5^{1/2}+d_1^{1/2}\gamma_{1}^{2} )\,,\quad e^{1}=\frac{(d_1d_5)^\frac14}{\sqrt2}\frac{\dd r}{r}\,, \quad e^{2}=\dd x_{+}-\frac{\sqrt{d_1}r^2}{2}\dd x_-(d_5^{1/2}-d_1^{1/2}\gamma_{1}^{2} )\,,\nonumber \\
        e^{3} & =\frac{(d_1d_5)^\frac14}{2\sqrt2}\dd\psi\,,\qquad e^{4}=\frac{(d_1d_5)^\frac14}{2\sqrt2}\sin\psi\dd\varphi_{1}\,,\qquad e^{5}=\frac{d_5^\frac34}{2\sqrt2d_1^\frac14}e^{\Phi}(\dd\varphi_{2}-\cos\psi\dd\varphi_{1})\,,\\
        e^{6} & =\left(\frac{d_1}{d_5}\right)^\frac14\dd y^{1}\,,\qquad e^{7}=\left(\frac{d_1}{d_5}\right)^\frac14\dd y^{2}\,,\qquad e^{8}=\left(\frac{d_5}{d_1}\right)^\frac14e^\Phi\dd y^{3}\,,\qquad e^{9}=\left(\frac{d_1}{d_5}\right)^\frac14\dd y^{4}\,,\nonumber 
\end{align}
the Killing spinor indeed reads
\begin{align}
    \epsilon(\psi,\varphi_1)&=e^{\frac{\psi}{2}\Gamma^{13}}e^{-\frac{\varphi_1}{2}\Gamma^{15}\sigma_1}P_1 P_2 P_3\epsilon_0,\nonumber\\
    P_{1} & =\frac{1}{2}(1-\Gamma^{02})\,,\nonumber \\
    P_{2} & =\frac{1}{2}\left[1+e^\Phi\left(-\sqrt{8d_5}\gamma_{2}\Gamma^{5}\sigma_{3}+\sqrt\frac{d_5}{d_1}\Gamma^{8}\right)\Gamma^{679}\right]\,,\nonumber \\
    P_{3} & =\frac{1}{2}(1+\sigma_{1}\Gamma^{6789})\,,
\end{align}
where $\epsilon_0$ is a constant spinor, as usual. The near-horizon configuration presented above provides an example of a supersymmetric double deformed solution, i.e. with two independent parameters, obtained through two TsT dualities. 

\subsubsection{A larger class of solutions}\label{alcos}
Since the near horizon configuration \eqref{firstTsTnearhorizon} remains solution to the six-dimensional half-maximal model in \cite{Maurelli:2025iba}, it can be generalised to make it covariant with respect to the residual ${\rm SO}(4)$ global symmetry, once we fix the ${\rm SO}(4,4)/{{\rm SO}(4)\times {\rm SO}(4)}$ scalars to the origin.  In carrying out this procedure, one quickly realises that additional solutions exist which differ only in their flux configurations and yet remain physically distinct, as they preserve different amounts of supersymmetry. Let us start by considering the following ansatz:
\begin{align}
\dd s^{2} & =c_0\left(\frac{\dd {r}^{2}}{r^2}-r^2 \dd x_-\dd x_{+}+\frac{1}{4}[\dd\psi^{2}+\sin^{2}\psi\dd\varphi_{1}^{2}]\right)- c_1\mathsf{A}_{1}^2+c_2\mathsf{A}_2^2+\dd y_1^2+\dd y_2^2+\dd y_3^2+\dd y_4^2\,,\nonumber\\
\mathsf{H}_3&=k_1\,\dd \mathsf{A}_1\wedge \vec{W}\cdot \dd\vec{y}+k_2\,\dd \mathsf{A}_2\wedge \vec{X}\cdot \dd\vec{y}\,, \qquad \Phi=p_0\,,\nonumber\\
\mathsf{F}_3&=p_1 r\, \dd r\wedge \dd x_-\wedge \dd x_+-\frac{p_2}{4}\,\sin\psi\,\dd\psi\wedge \dd\varphi_1\wedge \dd\varphi_2\,,\nonumber\\
\mathsf{F}_5&=G_5 + \star G_5\,,\nonumber\\
G_5&=r\, \dd r\wedge \dd x_-\wedge \dd x_+\wedge \mathsf{A}_2\wedge \vec{X}\cdot \dd\vec{y}-\frac{1}{4}\,\sin\psi\dd\psi\wedge \dd\varphi_1\wedge \dd\varphi_2\wedge \mathsf{A}_1\wedge \vec{W}\cdot \dd\vec{y}\,, \\
\star G_5 &= -\frac{1}{2 \sqrt{c_0 c_2}} (\dd \mathsf{A}_1 W^a + \dd \mathsf{A}_2 X^a) \frac{1}{3!} \epsilon_{a b c d} \wedge \dd y^{b}\wedge \dd y^{c}\wedge \dd y^{d}\,
\end{align}
where $\mathsf{A}_1 = r^2 \dd x_{-}\, , \, \mathsf{A}_2 = \dd \varphi_2 - \cos \psi \dd \varphi_1$. The vectors $\vec{W},\,\vec{X}$ are constant ${\rm SO}(4)$-vectors: by fixing the ${\rm SO}(4)$ symmetry of the torus, we can always bring such vectors to the following normal form:
$$\vec{W}=\{0,0,0,W_4\}\,,\,\,\,\vec{X}=\{0,0,X_3,\,X_4\}\,,$$ with a residual ${\rm SO}(2)$-symmetry. Let us distinguish the possible solutions, according to the scalar product $\vec X\cdot\vec W$:
\begin{itemize}
    \item $\vec X\cdot\vec W \geq 0$ 
    \begin{align}\hspace{-.7cm}
        k_1=-k_2\, \quad p_1=-p_2=\frac{1}{k_2}\,, \quad p_0=\frac{1}{2}\log(4 c_0 c_2 k_2^2)\,,\quad |\vec{W}|^2=\frac{c_1}{k_1^2}\,, \quad |\vec{X}|^2=\frac{c_0-4c_2}{4k_2^2}\,,
    \end{align}
    \item $\vec X\cdot\vec W=0$
    \begin{align}\hspace{-0.7cm}\label{8charges}
        k_1=k_2\,, \quad p_1= p_2=\frac{1}{k_2}\,, \quad p_0=\frac{1}{2}\log(4 c_0 c_2 k_2^2)\,,\quad |\vec{W}|^2=\frac{c_1}{k_2^2}\,, \quad |\vec{X}|^2=\frac{c_0-4c_2}{4k_2^2}\,.
    \end{align}
\end{itemize}
Both solutions depend on four free parameters, $c_0$, $c_1$, $c_2$ and $k_2$: the first one could in principle be removed by trombone symmetry, and it is related to the $\AdS$ radius, whereas $c_1$ and $c_2$ are responsible for deforming the  $\AdS_3$ and $\mathrm{S}^3$ geometries, respectively. Finally, the parameter $k_2$ determines the charge of $\mathsf F_3$. In both cases, the total metric belongs to the Killing-Kundt class. 

The first solution, which preserves four supercharges for any value of $\vec X\cdot \vec W$, corresponds to the $\mathrm{SO}(4)$-covariant generalisation of \eqref{firstTsTnearhorizon} for a vanishing value of such scalar product. The second solution, despite only differing by two signs, describes a truly independent configuration, as it preserves eight supercharges. Details on the derivation of the Killing spinor are given in Appendix \ref{proofsusy2}. 

\subsection{Supersymmetric Killing-Kundt solution with 6D Kaluza-Klein vector}
Let us now consider a different brane configuration, obtained through the following two subsequent TsT dualities 
$(x_+,\varphi_2,\gamma_1),$ $(\varphi_2,y_4,\gamma_2)$.
The solution is written in terms of the following $1$-forms
\begin{align}\label{nearhorizon1forms}
\mathbf{A}_{1} & =r^2\dd x_{-}\,,\nonumber\\
\mathbf{A}_{2} & =\dd\varphi_{2}-\cos\psi\dd\varphi_{1}\,
\end{align}
and it reads
\begin{align}\label{secondTsT}
\dd s^{2} & =\sqrt{H_{1}H_{5}}\left[\dd r^{2}+\frac{r^{2}}{4}\left(\dd\psi^{2}+\sin^{2}\psi\,\dd\varphi_{1}^{2}+\frac{H_{5}}{H_{1}}e^{2\Phi}\mathbf{A}_{2}^{2}\right)\right]-\frac{1}{r^2\sqrt{H_1 H_5}}(\mathbf{A}_{1} \dd x_++\frac{\gamma_1^2}{16}\mathbf{A}_{1}^2)\nonumber \\
& +\sqrt{\frac{H_{1}}{H_{5}}}\left(\dd y_{1}^{2}+\dd y_{2}^{2}+\dd y_{3}^{2}+\frac{H_{5}}{H_{1}}e^{2\Phi}\left(\dd y_4-\frac{\gamma_1\gamma_2}{8}\mathbf{A}_{1})^2\right)\right)\,,\nonumber\\
\mathsf B_{2} & =\frac{e^{2\Phi}H_5}{8H_1}(\gamma_1\mathbf{A}_{1}\wedge\mathbf{A}_{2}-2\gamma_2 r^2H_1 \mathbf{A}_{2}\wedge\dd y_4)\,,\qquad e^{2\Phi}=\frac{4H_{1}}{(4+r^{2}\gamma_{2}^{2}H_{1})H_{5}}\,,\nonumber\\
\mathsf{F}_{3} & =\frac{1}{8}(d_{5}\sin\psi\dd\varphi_{1}\wedge\dd\varphi_{2}\wedge\dd\psi+\frac{4d_{1}}{r^{3}H_{1}^{2}}\dd r\wedge\dd x_{-}\wedge\dd x_{+})\,,\qquad\mathsf{F}_{1}=0\,,\nonumber\\
\mathsf{F}_{5} & =e^{2\Phi}\frac{d_1\gamma_2H_5}{8r^3H_1^2}(\mathbf{A}_{1}\wedge\dd r\wedge \dd x_+\wedge\mathbf{A}_{2}\wedge\dd y_4)+\frac{d_1\gamma_2}{8}\sin\psi\dd \varphi_1\wedge \dd \psi\wedge\dd y_1\wedge \dd y_2\wedge \dd y_3\,.
\end{align}
The solution is everywhere regular, as all components of the Riemann tensor, in the rigid basis, are finite. The explicit appearance of $\mathbf{A}_2$ implies, as in the previous case, that the above expression is only valid in a single patch of $S^2$. Differently from the previous TsT generated solution, \eqref{secondTsT} features a Kaluza-Klein vector in $d=6$, proportional to $\mathbf{A}_{1}$: for this reason it cannot be described within the $d=6$, $\mathcal N=(1,1)$ consistent truncation considered in \cite{Maurelli:2025iba}. 

The total solution describes a D1-D5 system in the presence of NS5-branes and standard D3-branes, whose associated conserved charges are given by \eqref{charges first tst}, \eqref{chargeNS5}. Moreover, despite the presence of a Kaluza-Klein vector, the spacetime is again Killing-Kundt, with null vectors as in \eqref{nullvectorsKundt}. Finally, one can show that the solution is supersymmetric and that it preserves four supercharges. The explicit form of the Killing spinor can be derived by following the derivation in Appendix \ref{proofsusy}.

\paragraph{Near horizon geometry analysis}
Let us now study the near horizon region of the above solution, by performing, as in the previous case, the non-relativistic rescaling in \eqref{nearhorizonlimit}. The solution is again written in terms of the contact structures \eqref{nearhorizon1forms} as 
\begin{align}\label{nearhorizonsecondtst}
\dd s^{2} & =\frac{\sqrt{d_{1}d_{5}}}{2}\left(\frac{\dd {r}^{2}}{r^2}-4r^2 \dd x_-\dd x_{+}+\frac{1}{4}[\dd\psi^{2}+\sin^{2}\psi\dd\varphi_{1}^{2}+(\dd\varphi_{2}-\cos\psi\dd\varphi_{1})^{2}]\right)\nonumber\\
&+\frac{\sqrt{d_{1}d_{5}}}{8}\!\left(\!-d_1d_5 \gamma_1^2 \mathsf{A}_{1}^2+\!\left(e^{2\Phi}\frac{d_5}{d_1}-1\right)\!\mathsf{A}_2^2\right)\!+\!\sqrt{\frac{d_{1}}{d_{5}}}\!\left(\!\dd y_{1}^{2}+\dd y_{2}^{2}+\dd y_{3}^{2}+\frac{d_{5}}{d_{1}}e^{2\Phi}\left(\dd y_4-d_1 d_5\gamma_1\gamma_2\mathsf{A}_{1}\right)^2\!\right),\nonumber\\
\mathsf B_{2} & =\frac{e^{2\Phi}d_5}{8}(\gamma_1d_5\mathsf{A}_{1}\wedge\mathsf{A}_{2}-8\gamma_2 \mathsf{A}_{2}\wedge\dd y_4)\,,\qquad e^{2\Phi}=\frac{d_1}{d_5(1+8d_1\gamma_2^2)}\,,\nonumber\\
\mathsf{F}_{3} & =\frac{d_5}{8}(\sin\psi\dd\varphi_{1}\wedge\dd\varphi_{2}\wedge\dd\psi+16 r\dd r\wedge\dd x_{-}\wedge\dd x_{+})\,,\qquad\mathsf{F}_{1}=0\,,\nonumber\\
\mathsf{F}_{5} & =e^{2\Phi}\frac{2d_5^2\gamma_2}{r}(\mathsf{A}_{1}\wedge\dd r\wedge \dd x_+\wedge\mathsf{A}_{2}\wedge\dd y_4)+d_1\gamma_2\sin\psi\dd \varphi_1\wedge \dd \psi\wedge\dd y_1\wedge \dd y_2\wedge \dd y_3\,.
\end{align}
From a 6-dimensional point of view, the solution describes a $\mathrm{WAdS}_3\times \mathrm{WS}^3$ geometry, in the presence of a null Kaluza-Klein vector, two vectors coming from the Kalb-Ramond and $C_4$ fields, two 2-form potentials descending from $C_2$, $B_2$. This shows that the solution \eqref{secondTsT} interpolates between a locally flat spacetime in the asymptotic region and a space with geometry $\mathrm{WAdS}_3\times \mathrm{WS}^3\ltimes T^4$, where the torus is fibered over $\mathrm{WAdS}_3$. As in the previous case, the near-horizon limit does not feature a supersymmetry enhancement, as the configuration \eqref{nearhorizonsecondtst} still preserves four supercharges. In particular, given the vielbein
\begin{align}
        e^{0} & =\dd x_{+}\!+\!\frac{\sqrt{d_1 d_5}r^2}{2}\dd x_-\!\left(\!\frac{\sqrt{d_1 d_5}}{8}\gamma_{1}^{2}\!+\!1 \!\right), \!\!\quad e^{1}=\frac{(d_1d_5)^\frac14}{\sqrt2}\frac{\dd r}{r},\quad \! \! e^{2}=\dd x_{+}\!+\!\frac{\sqrt{d_1 d_5}r^2}{2}\dd x_-\!\left(\!\frac{\sqrt{d_1 d_5}}{8}\gamma_{1}^{2}\!-\!1\! \right),\nonumber \\
        e^{3} & =\frac{(d_1d_5)^\frac14}{2\sqrt2}\dd\psi\,,\qquad e^{4}=\frac{(d_1d_5)^\frac14}{2\sqrt2}\sin\psi\dd\varphi_{1}\,,\qquad e^{5}=\frac{d_5^\frac34}{2\sqrt2d_1^\frac14}e^{\Phi}(\dd\varphi_{2}-\cos\psi\dd\varphi_{1})\,,\\
        e^{6} & =\left(\frac{d_1}{d_5}\right)^\frac14\dd y^{1}\,,\quad e^{7}=\left(\frac{d_1}{d_5}\right)^\frac14\dd y^{2}\,,\quad e^{8}=\left(\frac{d_1}{d_5}\right)^\frac14\dd y^{3}\,,\quad e^{9}=\left(\frac{d_5}{d_1}\right)^\frac14e^\Phi(\dd y^{4}-d_1d_5\gamma_1\gamma_2\mathsf{A}_1)\,,\nonumber 
\end{align}
the Killing spinor reads
\begin{align}
    \epsilon(\psi,\varphi_1)&=e^{\frac{\psi}{2} \Gamma^{13}}e^{-\frac{\varphi_1}{2}\Gamma^{15}\sigma_1}P_1 P_2 P_3P_0\epsilon_0,\nonumber\\
    P_{1} & =\frac{1}{2}(1-\Gamma^{02})\,,\nonumber \\
    P_{2} & =\frac{1}{2}\left[1+e^\Phi\left(\sqrt{8d_5}\gamma_{2}\Gamma^{5}\sigma_{3}-\sqrt\frac{d_5}{d_1}\Gamma^{9}\right)\Gamma^{678}\right]\,,\nonumber \\
    P_{3} & =\frac{1}{2}(1+\sigma_{1}\Gamma^{6789})\,.
\end{align}

\section{Doubly-deformed black hole from TsT}\label{sec4}

The warped near-horizon geometries obtained so far, \eqref{firstTsTnearhorizon}, \eqref{nearhorizonsecondtst}, could be obtained by performing the corresponding TsT dualities directly on the $\AdS_3\times \mathrm{S}^3\times T^4$ solution. This indicates that, at least in the null case, the near-horizon limit commutes with the TsT dualities. The same cannot be claimed a priori for TsT's along spacelike coordinates, which in turn hinders the interpretation of spacelike $\mathrm{WAdS}_3$ spaces as resulting from the near-horizon limit of certain brane configurations. Nonetheless, TsT dualities along spacelike directions were essential to derive warped black hole configurations, free of closed timelike curves, from the $\AdS_3\times \mathrm{S}^3\times T^4$ solution \cite{Georgescu:2025jlx}. In this section, we will provide a generalisation of such black holes, featuring a warping of both $\mathrm{AdS}_3$ and $\mathrm S^3$, leaving the conjectured brane interpretation to a future endeavour. To this end, and to make contact with the notation of \cite{Georgescu:2025jlx}, let us rewrite the undeformed configuration as
\footnote{The configuration below is obtained from \eqref{firstTsTnearhorizon} by first setting $\gamma_1=\gamma_2=0$, $r\to \frac{r}{2}$, $d_1=d_5=2$, then performing the following change of coordinates
\begin{align}
    x_+ &= e^{2 T_u U}\sqrt{\frac{R-2 T_u T_v}{R+2 T_u T_v}}\,, \qquad x_-=- e^{2 T_v V}\sqrt{\frac{R-2 T_u T_v}{R+2 T_u T_v}}\,,\qquad r =e^{-T_u U- T_v V}\sqrt{\frac{R+2 T_u T_v}{4T_u T_v}}\,,
\end{align}
followed by an $\mathrm{SO}(2)$ rotation exchanging $\mathsf{F}_3$ and $\mathsf{H}_3$.
}
\begin{align}
    \dd s^2&=\frac{\dd R^{2}}{4\left(R^{2}-4T_{u}^{2}T_{v}^{2}\right)}+T_u^2 \dd U^2+R\dd U \dd V+\dd V^{2}T_{v}^{2}\nonumber\\
    &+\frac{1}{4}[\dd\psi^{2}+\sin^{2}\psi\dd\varphi_{1}^{2}+(\dd\varphi_{2}-\cos\psi\dd\varphi_{1})^{2}]+\dd\overset{\to}{y}{}^2\,,\nonumber\\
    \mathsf{H}_3&=\frac12 \dd R \wedge \dd U \wedge \dd V+\frac14 \sin\psi \dd\varphi_1\wedge \dd \varphi_2\wedge \dd\psi\,,\qquad \Phi=0\,,\nonumber\\
    \mathsf{F}_1&=\mathsf{F}_3=\mathsf{F}_5=0\,
\end{align}
Here $U,V$ are both spacelike coordinates, that describe, together with $R$, an $\AdS_3$ space, while $T_u$, $T_v$ are integration constants. Let us now consider the following double TsT duality $(y_4,V,T_v\sqrt{\Lambda-1}),$ $(y_3,\varphi_2,2\sqrt{\Delta-1})$. The total solution in the Einstein frame is rewritten as
\begin{align}
\dd s^{2} & =e^{-\frac{\Phi}{2}}\dd s^{2}(\mathrm{WAdS}_{3})+e^{-\frac{\Phi}{2}}\dd s^{2}(\mathrm{WS}^{3})+\left(\dd y_{4}+A_{1}\right)^{2}+\left(\dd y_{3}+A_{2}\right)^{2}+\dd y_{1}^{2}+\dd y_{2}^{2}\,,\nonumber\\
\mathsf{H}_{3} & =\frac{1}{2\Lambda}\dd R\wedge \dd U\wedge \dd V+\frac{1}{4\Delta}\sin\psi\dd\varphi_1\wedge\dd \varphi_2\wedge \dd \psi+e^{\frac{\Phi}{2}}\dd A_{1}\wedge\dd y_{4}+e^{\frac{\Phi}{2}}\dd A_{2}\wedge\dd y_{3}\,,\nonumber\\
e^{2\Phi} & =\frac{1}{\Lambda\Delta}\,,
\end{align}
with
\begin{align}
\dd s^{2}(\mathrm{WAdS}_{3})&=\frac{\dd R^{2}}{4\left(R^{2}-4T_{u}^{2}T_{v}^{2}\right)}+\frac{R\dd U}{\Lambda}\left[\dd V+\dd U\left(\frac{\Lambda T_{u}^{2}}{R}-\frac{\Lambda-1}{4T_{v}^{2}}R\right)\right]+\dd V^{2}\frac{T_{v}^{2}}{\Lambda}\,,\nonumber \\
\dd s^{2}(\mathrm{WS}^{3})&=\frac{1}{4}[\dd\psi^{2}+\sin^{2}\psi\dd\varphi_{1}^{2}+\Delta^{-1}(\dd\varphi_{2}-\cos\psi\dd\varphi_{1})^{2}]\,,\nonumber\\
A_{1} & =\frac{e^{-\frac{\Phi}{4}}}{2T_{v}}\sqrt{\frac{\Lambda-1}{\Lambda}}R\dd U\,,\qquad A_{2} =\frac{e^{-\frac{\Phi}{4}}}{2}\sqrt{\frac{\Delta-1}{\Delta}}\cos\psi\dd\varphi_{1}\,,
\end{align}
after a rescaling of the torus coordinates, $y_{1,2}\to e^{\Phi/4}y_{1,2}$, $y_3\to -\sqrt{\Delta}e^{\Phi/4}y_{3}$, $y_4\to \sqrt{\Lambda}e^{\Phi/4}y_{4}$. 

Black hole solutions are usually obtained through a suitable compactification of an asymptotically spacelike coordinate. However, warped black holes in superstring theory are rare, compared to three-dimensional, higher derivative gravity theories, like Topologically Massive Gravity. Indeed, the compactification of a spacelike coordinate usually introduces closed timelike curves \cite{Detournay:2012pc}. As already noticed in \cite{Georgescu:2025jlx}, the presence of a mixing term between $\mathrm{W}\AdS_3$ and the torus is the key ingredient that allows for consistent black holes. Indeed, let us consider the following change of coordinates 
\begin{align}
U=-T+ \varphi,\qquad V = T+\varphi\,.
\end{align}
The full solution is then described in terms of $T,\varphi,r,\psi,\varphi_1,\varphi_2,y_1,y_2,y_3,y_4$, where $T$ and $\varphi$ are timelike and spacelike at radial infinity, respectively. For this reason, the latter can be identified as $\varphi\sim\varphi+2\pi$, which corresponds in the $U,V$ coordinates to $U\sim U+2\pi$, $V\sim V+2\pi$. Such quotient introduces a causal singularity at $R = -T_u^2\Lambda-T_v^2$, where the Killing vector $K_\varphi$ would become null, while $R= R_H = 2 T_u T_v$ is the corresponding event horizon. 

We end this section by showing that the above black hole solution satisfies the first law of thermodynamics. To this end, let us compute the temperature $\mathsf{T}$ by means of the Bardeen-Hawking-Carter formula \cite{Bardeen:1973gs}. By defining
\begin{align}
\ell^\mu \partial_\mu & = \partial_T +  \frac{T_u-T_v}{T_u+T_v}\partial_\varphi + \frac{2 T_u T_v \sqrt{\Lambda-1}}{T_u+T_v} \frac{\Delta^{1/8}}{\Lambda^{3/8}} \partial_{y_4} \, , \\
n_\mu \dd x^\mu & = -\dd T - \frac{\sqrt{R^2+4 R T_v^2 + 4 T_v^4 -R^2 \Lambda + 4 T_u^2 T_v^2 \Lambda}}{4 T_v (r^2-4 T_u^2 T_v^2)} \dd R\,,
\end{align}
such that $\ell^\mu\ell_\mu|_{R=R_H} = 0 $, $n^\mu n_\mu=0$ and $\ell^\mu n_\mu= -1$, the temperature is given by $2 \pi \mathsf{T}=n^\mu\nabla_\mu \ell_\nu \ell^\nu$. The entropy can instead be computed in terms of the Bekenstein-Hawking formula. Temperature and entropy explicitly read
\begin{align*}
\mathsf{T} & =\frac{2T_{u}T_{v}}{\pi(T_{u}+T_{v})}\,,\\
\mathsf{S} & =\frac{\mathrm{Area}}{4G}=\frac{16\pi^{7}}{G}(T_{u}+T_{v})\,.
\end{align*}
The conserved charges associated with the Killing vectors can be computed in the covariant phase space formalism, see \cite{Compere:2019qed} for an extensive introduction. In particular, we have
\begin{align}
M&\equiv K[\partial_T]  =\frac{8\pi^6(T_{u}^{2}\Lambda+T_{v}^{2})}{G\Lambda}\,,\nonumber \\
J_{\varphi} &\equiv K[-\partial_\varphi] =\frac{8\pi^6(T_{u}^{2}\Lambda-T_{v}^{2})}{G\Lambda}\,,\\
J_{y_{4}} &\equiv K\left[-2\partial_{y_4}\right] =-\frac{16\pi^6T_{v}\sqrt{\Lambda-1}}{\Delta^{1/8}\Lambda^{5/8}}\,.\nonumber 
\end{align}
Finally, the angular velocity can be read from the components of $\ell$ as
\begin{align}
\Omega_{\varphi} & =\frac{T_u-T_v}{T_u+T_v}\,,\qquad\Omega_{y_{4}}=\frac{2 T_u T_v \sqrt{\Lambda-1}}{T_u+T_v} \frac{\Delta^{1/8}}{\Lambda^{3/8}}\,.
\end{align}
The quantities obtained above satisfy the first law of thermodynamics:
\begin{align}
\delta M & =\mathsf{T}\delta\mathsf{S}+\Omega_{\varphi}\delta J_{\varphi}+\Omega_{y_{4}}\delta J_{y_{4}}\,,
\end{align}
which can be integrated to give the following Smarr relation
\begin{align}
2M & =\mathsf{TS}+2\Omega_{\varphi}J_{\varphi}+\Omega_{y_{4}}J_{y_{4}}\,.
\end{align}

\section{Conclusions}\label{conclusions}

In this work, we constructed, by applying the TsT generating technique to the D1-D5 system, novel Type IIB supergravity backgrounds, interpolating between an asymptotically locally flat spacetime at radial infinity and null-${\rm WAdS}_3 \times {\rm WS}^3 \times T^4$ in the near-horizon region. A distinctive aspect of our construction is the choice of a null coordinate as one of the T-duality directions. Since null-${\rm WAdS}_3$ does not require a Hopf-fibration description and its contact structure is null, the TsT transformation can be implemented directly along this direction. This greatly simplifies the procedure and makes the construction technically natural. What is particularly interesting is that this choice continues to work smoothly for the full interpolating backgrounds, not only in the near-horizon limit. The latter, in contrast to the standard D1-D5 system, is intrinsically non-relativistic: this difference is not accidental, but deeply connected to the structure of the dual field theories. Indeed, the conjectured dual theories to warped ${\rm AdS}_3$ backgrounds are Warped Conformal Field Theories (WCFTs), which are non-relativistic in the precise sense that their symmetry algebra is not the (Virasoro)$\times$(Virasoro) algebra of a 2d CFT, but the (Virasoro)$\times \mathrm{U}(1)$-Kac-Moody algebra. The presence of only a single copy of the Virasoro algebra reflects the absence of full Lorentz invariance, and the U(1) Kac-Moody factor captures the preferred null direction along which the system is warped. The choice of a lightcone coordinate also plays a fundamental role in preserving supersymmetries, as the corresponding 1-forms $\mathbf{A}_{1}$, $\mathsf{A}_{1}$, proportional to $\dd x_-$ are both null, leading to simplifications of the corresponding integrability equations.

Overall, our results support the idea that warped ${\rm WAdS}_3 \times {\rm WS}^3 \times T_4$ backgrounds generally admit a brane construction, paving the way for a systematic characterization of their brane origins. What makes this task particularly challenging is the fact that, to generate spacelike and timelike warpings in the near-horizon, one needs to write $\AdS_3$ in the Hopf-fibration form. The change of coordinate between the Poincar\'e patch and the Hopf-fibration necessarily mixes all coordinates, which makes the harmonic functions $H_1$, $H_5$ depend on all $\AdS_3$, hindering the application of the TsT generating technique. Nonetheless, extending this level of understanding, particularly to the spacelike-warped case, would be highly desirable. Indeed, it would allow us to shed more light on the microscopic construction, within superstring theory, of the class of TsT-generated black hole solutions, of which we provided an example. These are the only consistent Type IIB supergravity warped black hole configurations known to the authors, see also \cite{Georgescu:2025jlx}. \par
It would also be interesting to understand if the whole family of supersymmetric solutions discussed in subsection \ref{alcos}, covariant with respect to the ${\rm O}(4)$ global symmetry group in the $D=6$ $\mathcal{N}=(1,1)$ consistent truncation, originate from a TsT transformation and to study the orbit of the 1/4 BPS solution \eqref{8charges}. This class of backgrounds is characterised by two ${\rm O}(4)$ vectors $\vec{X},\,\vec{W}$ and also contains, besides the solutions discussed earlier, for which the condition $\vec{X}\cdot \vec{W}=0$ holds, solutions in which the two constant vectors are parallel \cite{Maurelli:2025iba}, whose possible origin as the result of a TsT transformation of the undeformed background $\vec{X}=\vec{W}=\vec{0}$ is not clear.

Finally, in doubly-warped ${\rm WAdS}_3 \times {\rm WS}^3 \times T_4$ backgrounds with spacelike or timelike warping, supersymmetry is broken by continuous deformations and is restored if the warping on either factor is set to zero \cite{Maurelli:2025iba}. This suggests that, though non-supersymmetric, these backgrounds may still be perturbatively stable, at least within a certain range of the two continuous parameters. Assessing this stability issue is the subject of an ongoing investigation \cite{NEWPAPER}.

\section*{Aknowledgements}
We would like to thank Henning Sambtleben, Domenico Orlando and Ricardo Stuardo for useful discussions and comments.
R.N. was supported by the European Union and the Czech Ministry of Education, Youth and Sports
(Project: MSCA Fellowship CZ FZU III - CZ.02.01.01/00/22 010/0008598).

\appendix

\section{Conventions on Type IIB supergravity}\label{appIIB}

Throughout the text, we will follow the conventions presented in \cite{Tomasiello:2022dwe}. We will use Greek letters to denote ten-dimensional curved indices, while the latin alphabet will label rigid indices. The signature will be mostly plus.

For any $p$-form in 10 dimensions $\mathsf{F}_{p}$, we define
\begin{align}
|\mathsf{F}_{p}|^{2} & \equiv\frac{1}{p!}\mathsf{F}_{\mu_{1}\dots \mu_{p}}\mathsf{F}^{\mu_{1}\dots \mu_{p}}\,,\hspace{1.8cm}|\mathsf{F}_{p}|_{\mu\nu}^{2}\equiv\frac{1}{(p-1)!}\mathsf{F}_{\mu\nu_{1}\dots \nu_{p-1}}\mathsf{F}_{\nu}{}^{\nu_{1}\dots \nu_{p-1}}\,,\label{squareform}\\
\star\mathsf{F}_{p} & \equiv\frac{\sqrt{-g}}{p!(D-p)!}\dd x^{\mu_{1}\dots \mu_{D-p}}\epsilon_{\mu_{1}\dots \mu_{D-p}\nu_{1}\dots \nu_{p}}\mathsf{F}^{\nu_{1}\dots \nu_{p}}\,,
\end{align}
with $\epsilon_{12\dots D}=-\epsilon^{12\dots D}=1$. Moreover, we define the 10-dimensional volume element as $\dd x^{\mu_0\ldots \mu_9}=-\epsilon^{\mu_0\ldots \mu_9}\dd^{10}x$.

In these conventions, the bosonic pseudo-action of Type IIB supergravity in the string frame reads
\begin{align}
I_{\mathrm{IIB}} & =\frac{1}{2\kappa^{2}}\int\dd^{10}x\sqrt{-g}\left[e^{-2\Phi}\left(R+4\partial_{\mu}\Phi\partial^{\mu}\Phi-\frac{1}{2}|\mathsf{H}_{3}|^{2}\right)-\frac{1}{2}|\mathsf{F}_{1}|^{2}-\frac{1}{2}|\mathsf{F}_{3}|^{2}-\frac{1}{4}|\mathsf{F}_{5}|^{2}\right]+\nonumber \\
 & \quad-\frac{1}{4\kappa^{2}}\int C_{4}\wedge \mathsf F_3\wedge \mathsf{H}_{3}\,,
\end{align}
where 
\begin{align}\label{FDefs}
\mathsf{F}_{1} & =\dd C_{0}\,,\hspace{1.2cm}\mathsf{F}_{3}=\dd C_{2}-C_{0}\wedge \mathsf{H}_{3}\,,\hspace{1.2cm}\mathsf{F}_{5}=\dd C_{4}-C_{2}\wedge \mathsf{H}_{3}\,,
\end{align}
and $\mathsf{F}_{5}$ is required to be self-dual:
$$\mathsf{F}_{5}=\star\mathsf{F}_{5}\,.$$
The bosonic equations of motion are
\begin{align}
0 & = \dd  \star \mathsf{F}_{1} + \mathsf{H}_3 \wedge \star \mathsf{F}_3 \,, \qquad 
0 = \dd \star \mathsf{F}_{3}+\mathsf{H}_{3}\wedge\mathsf{F}_{5} \,, \qquad 
0= \dd\mathsf{F}_{5}-\mathsf{H}_{3}\wedge\mathsf{F}_{3}\,, \nonumber\\
0 & = \dd(e^{-2\Phi}\star \mathsf{H}_{3})+\mathsf{F}_{5}\wedge\mathsf{F}_{3}-\mathsf{F}_{1}\wedge\star \mathsf{F}_{3}\,, \nonumber\\ 
0 & = 2R-|\mathsf{H}_{3}|^{2}-8e^{\Phi} \nabla^{\mu} \nabla_{\mu} e^{-\Phi}\,,\\
0 & = R_{\mu \nu}+2\nabla_{\mu }\nabla_{\nu}\Phi-\frac{1}{2}|\mathsf{H}_{3}|_{\mu \nu}^{2} -\frac{e^{2\Phi}}{2}\left[|\mathsf{F}_{1}|_{\mu \nu}^{2}+|\mathsf{F}_{3}|_{\mu \nu}^{2}+\frac{1}{2}|\mathsf{F}_{5}|_{\mu \nu}^{2}-\frac{1}{2}g_{\mu \nu}(|\mathsf{F}_{1}|^{2}+|\mathsf{F}_{3}|^{2})\right]\,. \nonumber
\end{align}

The fermionic fields of Type IIB supergravity are given by two gravitini $\boldsymbol{\Psi}_\mu=(\psi_\mu^{\alpha A})$ and two spin 1/2 fields $\boldsymbol{\lambda}=(\lambda^{\alpha A})$, with $\alpha=1,\ldots 32$ and $A=1,2$ a $\mathrm{SO}(2)$ doublet index. They are both Majorana-Weyl spinors, that is they satisfy
\begin{align}
   \Gamma_* \otimes\mathsf 1_2\boldsymbol{\Psi}_\mu&=\boldsymbol{\Psi}_\mu\,,\qquad \bar{\boldsymbol{\Psi}}_\mu=\boldsymbol{\Psi}_\mu^T  C\otimes\mathsf 1_2\,,\nonumber\\
    \Gamma_*\otimes\mathsf 1_2\boldsymbol{\lambda}&=-\boldsymbol{\lambda}\,,\qquad \bar{\boldsymbol{\lambda}}=\boldsymbol{\lambda}^T  C\otimes\mathsf 1_2\,. \label{pseudo-majorana condition}
\end{align}
Here $\Gamma_*$ is the chirality matrix, defined in terms of the Dirac matrices $\Gamma^{a}$ as $\Gamma_*=\Gamma^{0\ldots9}$, whereas $\mathds 1_2$ acts on the suppressed R-symmetry index.

The supersymmetry variations of the spinorial fields, expressed in terms of a fermionic parameter $\boldsymbol{\epsilon}$, are given by 
\begin{align}\label{deltalambda}
\delta\boldsymbol{\lambda} & =\not{\dd}\Phi\boldsymbol{\epsilon}-\frac{1}{2}\not{\mathsf{H}}_{3}\sigma_{3}\boldsymbol{\epsilon}-e^{\Phi}(\mathsf{F}_{1}\ri\sigma_{2}+\frac{1}{2}\mathsf{F}_{3}\sigma_{1})_{/}P_0\boldsymbol{\epsilon}\,,\\
\delta\boldsymbol{\Psi}_{\mu} & =\partial_{\mu}\boldsymbol{\epsilon}+\boldsymbol{W}_{\mu}\boldsymbol{\epsilon}\,,\label{Killingeq}\\
\boldsymbol{W}_{\mu} & =\frac14\omega^{ab}{}_\mu \Gamma_{ab}\boldsymbol{\epsilon}-\frac{1}{4\cdot2}\mathsf{H}_{\mu\rho\sigma}\Gamma^{\rho\sigma}\sigma_{3}\boldsymbol{\epsilon}+\frac{e^{\Phi}}{8}(\mathsf{F}_{1}\ri\sigma_{2}+\mathsf{F}_{3}\sigma_{1}+\frac{1}{2}\mathsf{F}_{5}\ri\sigma_{2})_{/}\Gamma_{\mu}P_0\boldsymbol{\epsilon}\,,
\end{align}
where $\sigma_{1,2,3}$ are the Pauli matrices, $P_0=\frac12(\mathsf{1}_{32}+\Gamma_*)\otimes \mathsf 1_2$ and the slash operation is defined for any $p$-form $\mathsf{F}_{p}$ as 
\begin{align}\label{slashed}
\not{\mathsf{F}}_{p}=(\mathsf{F}_{p})_{/} & \equiv\frac{1}{p!}F_{\mu_{1}\dots \mu_{p}}\hat\Gamma^{\mu_{1}\dots \mu_{p}}\,.
\end{align}

\section{The general TsT transformation}\label{appTsT}
Let us consider a solution of Type II supergravity, with $d$ cyclic coordinates $x^i$, i.e. coordinates associated with a translational isometry $x^i\rightarrow x^i+\xi^i$. Denoting by $x^\alpha$ the remaining coordinates, we can write the metric in the following general form 
\begin{align}
\dd s^{2} & =g_{\alpha\beta}\,\dd x^{\alpha}\dd x^{\beta}+G_{ij}(\dd x^{i}+A^{i})(\dd x^{j}+A^{j})\,.
\end{align}
We choose the cyclic coordinates $x^i$ so that the metric and the Kalb-Ramond field have no off-diagonal components $g_{i\alpha},\,B_{i\alpha}$, namely so that $A^i=A^i_\alpha\,{\rm d} x^\alpha=B_{i\alpha}=0$.
\subsection{Transformation of NS-NS fields}
Let us consider the O($d,d$) transformation on the worldsheet theory \cite{Giveon:1994fu}, acting on the pseudo-orthogonal vector $(x_M)\equiv (\tilde{x}_i,\, x^i)$ through the following matrix:
\begin{align}
\boldsymbol{\Lambda}\in\mathrm{O}(d,d)&\,,\qquad\boldsymbol{\Lambda}=\left(\begin{array}{cc}
A & B\\
C & D
\end{array}\right)\equiv\,(\Lambda_{M}{}^{N})\,\,:\,\,\,\,\boldsymbol{\Lambda}^T\cdot J\cdot \boldsymbol{\Lambda}=J\,,\nonumber \\
A&=(a_{i}{}^{j})\,,\qquad B=(b_{ij})\,,\qquad C=(c^{ij})\,,\qquad D=(d^{i}{}_{j})\,,
\end{align}
where $J\equiv\left(\begin{array}{cc}
0 & \mathds{1}\\
\mathds{1} & 0
\end{array}\right)$ is the O($d,d$)-invariant matrix. The corresponding action on the metric and Kalb-Ramond moduli $G_{ij},\,B_{ij}$ is conveniently described in terms of the following matrix
\begin{align}
E_{ij} & =G_{ij}+B_{ij}\,,
\end{align}
with ${\bf G}\equiv (G_{ij})$ being the internal metric in the string frame. The action of $\boldsymbol{\Lambda}$ on ${\bf E}\equiv (E_{ij})$ is
\begin{align}
{\bf E}\to {\bf E}^{\prime}=(A{\bf E}+B)\,(C{\bf E}+D)^{-1}\,,
\end{align}
from which we deduce the transformation property of the metric:
\begin{align}
    {\bf G}^{\prime}=(C{\bf E}+D)^{-T}\, {\bf G}\,(C{\bf E}+D)^{-1}\,.
\end{align}
The dilaton transforms as follows:
\begin{align}
e^{2\Phi^{\prime}} & =\frac{e^{2\Phi}}{\det(C\,{\bf E}+D)}\,,
\end{align}
so that we can define a O($d,d$)-invariant dilaton as
\begin{align}
\hat{\Phi} & \equiv\Phi-\frac{1}{4}\log(\det {\bf G})\,.
\end{align}
The TsT transformation, denoted by $(x^1,x^2,\gamma)$ involves $d=2$ cyclic directions $x^1,x^2$ and is implemented by the  element of O($2,2$) 
which results from the following consecutive transformations:
\begin{align}
x^1\rightarrow \tilde{x}_1&\,\,,\,\,\,\,\,\,\,T=\left(\begin{matrix}0 & 0 & 1 & 0\cr
0 & 1 & 0 & 0\cr
1 & 0 & 0 & 0\cr 0 & 0 & 0 & 1\end{matrix}\right)\,,\nonumber\\
x^2\rightarrow x^2-\gamma\,\tilde{x}_1&\,\,,\,\,\,\,\,\,\,s=\left(\begin{matrix}1 & \gamma & 0 & 0\cr
0 & 1 & 0 & 0\cr
0 & 0 & 1 & 0\cr 0 & 0 & -\gamma & 1\end{matrix}\right)\,,\nonumber\\
\tilde{x}_1\rightarrow x^{\prime 1}&\,\,,\,\,\,\,\,\,\,T=\left(\begin{matrix}0 & 0 & 1 & 0\cr
0 & 1 & 0 & 0\cr
1 & 0 & 0 & 0\cr 0 & 0 & 0 & 1\end{matrix}\right)\,
\end{align}
The total transformation explicitly reads
\begin{align}
\boldsymbol{\Lambda}_{TsT} &= T\cdot s\cdot T=\left(\begin{array}{cc}
\mathds{1} & 0\\
\boldsymbol{\gamma} & \mathds{1}
\end{array}\right)\,,\qquad\boldsymbol{\gamma}=(\gamma^{ij})=\left(\begin{matrix}0 & \gamma \cr -\gamma & 0\end{matrix}\right)\,.\label{TsTtra}
\end{align}

\subsection{Transformation of RR fields}
Let us now describe the action of an SO($d,d$)-transformation on RR fields. We shall mainly follow the review \cite{Fukuma:1999jt}.\footnote{Our notation differs from that of \cite{Fukuma:1999jt} in the sign of the Kalb-Ramond 2-form $B_2$.}

Let us start by constructing the spinorial representation of $\mathrm{Spin}(d,d)$: to this end let $C_{p}$ be the RR $p$-form potentials and let us define
\begin{align*}
\mathrm{D}_{0} & =C_{0}\,,\\
\mathrm{D}_{1} & =C_{1},,\\
\mathrm{D}_{2} & =C_{2}-\mathsf{B}_{2}\wedge C_{0}\,,\\
\mathrm{D}_{3} & =C_{3}-\mathsf{B}_{2}\wedge C_{1}\,,\\
\mathrm{D}_{4} & =C_{4}-\frac{1}{2}\mathsf{B}_{2}\wedge C_{2}+\frac{1}{2}\mathsf{B}_{2}\wedge \mathsf{B}_{2}\wedge C_{0}\,,
\end{align*}
whose field strengths are 
\begin{align*}
\mathsf{F}_{1} & =\dd\mathsf{D}_{0}\,,\\
\mathsf{F}_{2} & =\dd\mathsf{D}_{1}\,,\\
\mathsf{F}_{3} & =\dd\mathsf{D}_{2}+\mathsf{B}_{2}\wedge\dd\mathsf{D}_{0}\,,\\
\mathsf{F}_{4} & =\dd\mathsf{D}_{3}+\mathsf{B}_{2}\wedge\dd\mathsf{D}_{1}\,,\\
\mathsf{F}_{5} & =\dd\mathsf{D}_{4}+\mathsf{B}_{2}\wedge\dd\mathsf{D}_{2}+\frac{1}{2}\mathsf{B}_{2}\wedge \mathsf{B}_{2}\wedge\dd\mathsf{D}_{0}\,.
\end{align*}
The Bianchi identities satisfied by the above field strengths are $\dd \mathsf{F}_{k}=\mathsf{H}_{3}\wedge \mathsf{F}_{k-2}$. Extending the definition of the field strengths with the Hodge dual
\begin{align*}
\mathsf{F}_{9} & =\star \mathsf{F}_{1}\,,\qquad \mathsf{F}_{8}=-\star \mathsf{F}_{2}\,,\\
\mathsf{F}_{7} & =-\star \mathsf{F}_{3}\,,\qquad \mathsf{F}_{6}=\star \mathsf{F}_{4}\,,\\
\mathsf{F}_{5} & =\star \mathsf{F}_{5}\,,
\end{align*}
we define the polyforms
\begin{align}
\mathsf{D} & =\sum_{p=0}^{8}\mathsf{D}_{p}\,,\qquad \mathsf{F}=\sum_{k=1}^{9}\mathsf{F}_{k}\,.
\end{align}
We can show that
\begin{align}
\mathsf{F} & =e^{\mathsf{B}}\wedge\dd\mathsf{D}\implies\dd \mathsf{F}=\mathsf{H}_{3}\wedge \mathsf{F}\,.
\end{align}
Under the action of $\mathrm{Spin}(d,d)$, both $\mathrm{D}$ and $\dd\mathrm{D}$ transform
in the spinorial representation, which can be constructed using creation/annihilation
operators $\psi^{\dagger i},\psi_{i}$ of $d$ fermions harmonic oscillators
\begin{align}
\{\psi_{i},\psi^{\dagger j}\} & =\delta_{i}^{j}\,,\qquad\{\psi^{\dagger i},\psi^{\dagger j}\}=\{\psi_{i},\psi_{j}\}=0\,.
\end{align}
The spinorial representation is described by the vectors
\begin{align}
|\alpha\rangle & =\psi^{\dagger i_{1}}\cdots\psi^{\dagger i_{\alpha}}|0\rangle\,,
\end{align}
with the condition $\psi_{i}|0\rangle=0\,,\forall i=1,\dots,d$. By defining 
\begin{align}
\Psi^{M} & =\left(\begin{array}{c}
\psi^{\dagger i}\\
\psi_{i}
\end{array}\right)\,,\qquad\{\Psi^{M},\Psi^{N}\}=J^{MN}\mathds{1}\,,
\end{align}
the spinorial representation of a generator $T_{MN}=-T_{NM}$ of $\mathrm{Spin}(d,d)$
can be written as
\begin{align}
S(T_{MN}) & =\frac{1}{2}T_{IJ|KL}\Psi^{K}\Psi^{L}=\frac{1}{2}(J_{IK}J_{JL}-J_{JK}J_{IL})\Psi^{K}\Psi^{L}=J_{MP}J_{NQ}\Psi^{P}\Psi^{Q}\,,\label{spinodd}
\end{align}
where we have used the fundamental representation of the O($d,d$)-generators $T_{MN}$:
\begin{align}
(T_{IJ})_{K}{}^{L} & =J_{IK}\delta_{J}^{L}-J_{JK}\delta_{I}^{L}\,.
\end{align}
One can verify that the \eqref{spinodd} generators close indeed the O($d,d$)-commutation relations:
\begin{align}
[S(T_{I_{1}I_{2}}),S(T_{I_{3}I_{4}})] & =J_{I_{1}I_{4}}S(T_{I_{2}I_{3}})+J_{I_{2}I_{3}}S(T_{I_{1}I_{4}})-J_{I_{1}I_{3}}S(T_{I_{2}I_{4}})-J_{I_{2}I_{4}}S(T_{I_{1}I_{3}})\,,
\end{align}
by using the identity
\begin{align}
[AB,CD] & =A\{B,C\}D-AC\{B,D\}-C\{A,D\}B+\{A,C\}DB\,,
\end{align}
 $A,B,C,D$ being fermionic operators.\footnote{
One can also verify that $\Psi_M$ transforms as a pseudo-orthogonal vector
\begin{align}
[S(T_{IJ}),\Psi_{K}] & =\frac{1}{2}T_{IJ|PQ}[\Psi^{P}\Psi^{Q},\Psi_{K}]= -T_{IJ|K}{}^P\Psi_P\,.
\end{align}}
Due to the isomorphism between the space of spinors and that of differential forms, that is
\begin{align}
|i_{1}\dots i_{k}\rangle & =\psi^{\dagger i_{1}}\dots\psi^{\dagger i_{k}}|0\rangle\,\leftrightarrow\,\dd x^{i_{1}}\wedge\dots\wedge\dd x^{i_{k}}\,,
\end{align}
the action of $\psi^{\dagger i}$ and $\psi_{i}$ is equivalent to the wedge and contraction operation, respectively
\begin{align}
\psi^{\dagger i}|i_{1}\dots i_{k}\rangle & &=|ii_{1}\dots i_{k}\rangle\,\leftrightarrow\,\dd x^{i}\wedge(\dd x^{i_{1}}\wedge\dots\wedge\dd x^{i_{k}})\,,\\
\psi_{i}|i_{1}\dots i_{k}\rangle & &=k\delta_{i[i_{1}}|i_{2}\dots i_{k]}\rangle\,\leftrightarrow\,\iota_{x^{i}}(\dd x^{i_{1}}\wedge\dots\wedge\dd x^{i_{k}})\,.
\end{align}
A polyform $\mathsf{D}$ can then be described as an O($d,d$)-spinor transforming as follows:
\begin{align}
    \boldsymbol{\Lambda}=e^{\frac{1}{2}\,\lambda^{MN}S(J_{MN})}:\,\,\mathsf{D}\,\rightarrow\,\,\mathsf{D}'=  \boldsymbol{\Lambda}\cdot \mathsf{D}\,.\label{Dtras}
\end{align}
\paragraph{TsT transformation of RR fields}

The transformation law of the RR  fields under the action of an O($d,d$) element is then derived by extracting the $\dd \mathsf{D}$ polyform from the RR field strength polyform: 
\begin{align*}
\dd\mathsf{D}=e^{-\mathsf{B}_{2}}\,\mathsf{F}\,,
\end{align*}
and transforming $\dd \mathsf{D}$  under \eqref{Dtras}.\par
In the case of a TsT transformation, $d=2$ and the ${\rm O}(2,2)$ transformation is given in \eqref{TsTtra}, so that:
\begin{align}
    \boldsymbol{\Lambda}_{{\rm TsT}}=e^{\frac{1}{2}\,S(T_{ij})\gamma^{ij}}\,&:\,\,\,\dd\mathsf{D}\mapsto\dd\mathsf{D}^{\prime}=e^{\frac{1}{2}\,S(T_{ij})\gamma^{ij}}\cdot \dd\mathsf{D}\,,\nonumber\\
e^{\frac{1}{2}\,S(T_{ij})\gamma^{ij}}&=e^{\frac{1}{2}\,\gamma^{ij}\,\iota_i\wedge \iota_j}=1+\gamma\,\iota_1\wedge \iota_2\,.
\end{align}
The transformed field strength polyform is then given by
\begin{align}\mathsf{F}^{\prime}=e^{\mathsf{B}^{\prime}}\dd\mathsf{D}^{\prime}\,.
\end{align}

\section{Proof of equation \eqref{KSfirsttst}}\label{proofsusy}
In order to prove supersymmetry of such background, one needs to solve the Killing equations \eqref{Killingeq}. Integrability implies the following set of equations
\begin{align}
    0 &= (1-P_1)\epsilon\,,\nonumber\\
    0 &= (1-P_2)\epsilon\,,\nonumber\\
    0 &= (1-P_{3})\epsilon =\frac{1}{2}\left(1+\mathsf{x}(r)\mathbf{\Gamma}_{\mathtt{1}}+\mathsf{y}(r)\boldsymbol{\Gamma}_{\mathtt{2}}\right)\epsilon\,,\label{matricial1}
\end{align}
where
\begin{align}
\boldsymbol{\Gamma}_{\mathtt{1}} & =\sigma_{1}\,,\qquad\mathbf{\Gamma}_{\mathtt{2}}=\Gamma^{58}\ri\sigma_{2}\,,\qquad \mathsf{x}(r) =-\frac{2\sqrt{2}}{\sqrt{8+d_{1}\gamma_{2}^{2}+2r^{2}\gamma_{2}^{2}}}\,, \qquad \mathsf{y}(r)  =\frac{\sqrt{2r^2+d_{1}}\gamma_{2}}{\sqrt{8+d_{1}\gamma_{2}^{2}+2r^{2}\gamma_{2}^{2}}}\,.\nonumber
\end{align}
The Killing spinor equations then reduce to 
\begin{align}
    0 &=\partial_r\epsilon+W_r \epsilon\,,\\
    0 &= \partial_\psi \epsilon+W_\psi\epsilon\,,\\
    0 &= \partial_{\varphi_1}\epsilon+W_{\varphi_1}\epsilon\,,
\end{align}
where
\begin{align}
    W_{r} & =-\frac{8r}{(d_{1}+2r^{2})(8+d_{1}\gamma_{2}^{2}+2r^{2}\gamma_{2}^{2})}+\frac{2\sqrt{2}r}{(d_{1}+2r^{2})\sqrt{8+d_{1}\gamma_{2}^{2}+2r^{2}\gamma_{2}^{2}}}\sigma_{1}\,,\\
    W_\psi&=-\frac12 \Gamma^{13}\,,\label{eqpsi}\\
    W_{\varphi_1}&=\frac12\Gamma^{15}\sigma_1(\cos\psi+\Gamma^{45}\sigma_1 \sin\psi)=\frac12\Gamma^{15}\sigma_1(\cos\psi-\Gamma^{13} \sin\psi)=\frac12\Gamma^{15}\sigma_1e^{-\psi\Gamma^{13}}\,,\label{eqphi1}
\end{align}
thanks to equation \eqref{matricial1}. Equations \eqref{eqpsi}, \eqref{eqphi1} are solved by
\begin{align}
    \epsilon=e^{\frac{\psi}{2}\Gamma^{13}}e^{-\frac{\varphi_1}{2}\Gamma^{15}\sigma_1}\eta(r)\,.
\end{align}
This leaves the equation for the radial dependence, which can be rewritten as
\begin{align}
    \eta^{\prime}=[\mathsf{a}(r)+\mathsf{b}(r)\mathbf{\Gamma}_{\mathtt{1}}+\mathsf{c}(r)\mathbf{\Gamma}_{\mathtt{2}}]\eta\label{differential2}
\end{align}
with 
\begin{align}
    \mathsf{a}(r) & =\frac{8r}{(d_{1}+2r^{2})(8+d_{1}\gamma_{2}^{2}+2r^{2}\gamma_{2}^{2})}\,,\nonumber \\
    \mathsf{b}(r) & =-\frac{2\sqrt{2}r}{(d_{1}+2r^{2})\sqrt{8+d_{1}\gamma_{2}^{2}+2r^{2}\gamma_{2}^{2}}}\,.
\end{align}
Since the following equations are satisfied,
\begin{align}
    \mathsf{x}^2+\mathsf{y}^2=1\,, \qquad \frac{\dd\mathsf{x}}{\dd r}+2\mathsf{b}\mathsf{y}^{2} & =0\,,
\end{align}
the solution for the radial dependence can be obtained following \cite{Romans:1991nq} and is given by 
\begin{align}
\eta & =\frac12\left(\frac{d_{1}+2r^{2}}{8+(d_{1}+2r^{2})\gamma_{2}^{2}}\right)^{1/4}\left(\alpha_{+}-\alpha_{-}\mathbf{\Gamma}_{\mathtt{2}}\right)(1-\mathbf{\Gamma}_{\mathtt{1}})P_{1}P_{2}\epsilon_0\,,\\
\alpha_{\pm} & =\sqrt{\frac{1\pm\mathsf{x}(r)}{\mathsf{y}(r)}}\,,
\end{align}
where $\epsilon_0$ is a constant spinor. The full solution then reads
\begin{align}
    \epsilon=\frac12\left(\frac{d_{1}+2r^{2}}{8+(d_{1}+2r^{2})\gamma_{2}^{2}}\right)^{1/4}e^{\frac{\psi}{2}\Gamma^{13}}e^{-\frac{\varphi_1}{2}\Gamma^{15}\sigma_1}\left(\alpha_{+}-\alpha_{-}\mathbf{\Gamma}_{\mathtt{2}}\right)(1-\mathbf{\Gamma}_{\mathtt{1}})P_{1}P_{2}P_0\epsilon_0\,.
\end{align}

\section{Solving the Killing spinor equation for \eqref{8charges} background}\label{proofsusy2}
Let us consider the following vielbein basis
\begin{align}
        e^{0} & =\dd x_{+}+\frac{r^2}{4c_0}\dd x_-(4c_1+c_0^2)\,,\quad e^{1}=\sqrt{c_0}\frac{\dd r}{r}\,, \quad e^{2}=\dd x_{+}+\frac{r^2}{4c_0}\dd x_-(4c_1-c_0^2)\,,\nonumber \\
        e^{3} & =\frac{\sqrt{c_0}}{2}\dd\psi\,,\qquad e^{4}=\frac{\sqrt{c_0}}{2}\sin\psi\dd\varphi_{1}\,,\qquad e^{5}=\sqrt{c_2}(\dd\varphi_{2}-\cos\psi\dd\varphi_{1})\,,\\
        e^{6} & =\dd y^{1}\,,\qquad e^{7}=\dd y^{2}\,,\qquad e^{8}=\dd y^{3}\,,\qquad e^{9}=\dd y^{4}\,.\nonumber 
\end{align}
The integrability equations 
\begin{align}
    (\dd \boldsymbol{W}+\boldsymbol{W}\wedge \boldsymbol{W})\epsilon=0\,,
\end{align}
together with \eqref{deltalambda}, are solved by the following projectors
\begin{align}
P_1&=\frac{1}{2}\left(1+\frac{k_2}{\sqrt{c_0c_2}}(-\Gamma^0+\Gamma^2)\vec{W}\cdot\vec{\Gamma}\otimes \ri\sigma_2-\frac{k_2}{\sqrt{c_2}}(\vec{X}\cdot\vec{\Gamma})\Gamma^5\otimes\sigma_3-\frac12\sqrt\frac{c_0}{c_2}\sigma_1\right)\,,\\
P_{2} & =\frac{1}{2}\left(1+\Gamma^{6789}\sigma_{1}\right)\,,
\end{align}
where $\vec{\Gamma}=\{\Gamma^6,\Gamma^7,\Gamma^8,\Gamma^9\}$.
The non-zero components of $W$ are given by
\begin{align}
    W_\psi&=\sqrt{\frac{c_2}{c_0}}\Gamma^{45}-\frac{k_2}{\sqrt{c_0}}\Gamma^4(\vec{X}\cdot\vec{\Gamma})\otimes\sigma_3\,,\nonumber\\
    W_{\varphi_1}&=-\frac12\cos\psi\Gamma^{34}-\sqrt{\frac{c_2}{c_0}}\sin\psi\Gamma^{35}+\frac{k_2}{\sqrt{c_0}}\sin\psi\Gamma^{3}(\vec{X}\cdot \vec{\Gamma})\otimes \sigma_3\,,\nonumber\\
    W_{r}&=\frac1r\left(-\Gamma^{02}+\frac{2k_2}{c_0}(\Gamma^2-\Gamma^0)(\vec{W}\cdot\vec{\Gamma})\otimes \sigma_3\right)\equiv\frac1r w_r\,,\nonumber\\
    W_{x_-}&=r^2 \Gamma^{01}\bigg[a_+\!+\!a_-\Gamma^{02}\!+\!(b_++b_-\Gamma^{02})\Gamma^5\vec{X}\cdot\vec{\Gamma}\otimes \ri\sigma_2\!+\!c(3\Gamma^0+\Gamma^{2})\vec{W}\cdot\vec \Gamma\otimes \sigma_3\!+\!(d_++d_-\Gamma^{02})\otimes\sigma_1\bigg]\nonumber\\
    &\equiv r^2 w_{x_-}\,,
\end{align}
with 
\begin{align}
    a_\pm=-\frac{20c_1\pm3c_0^2}{16 c_0^{\frac32}},\qquad b_\pm=\pm\frac{(c_0^2\mp 4c_1)k_2}{8c_0^2}, \qquad c=\frac{k_2}{4\sqrt{c_0}},\qquad d_\pm=\pm \frac{(c_0^2\mp 4c_1)\sqrt{c_2}}{8c_0^2}\,.
\end{align}
The $r$, $\psi$, $\varphi_1$ components of the Killing Spinor equation are solved by 
\begin{align}
    \epsilon=e^{-\psi W_\psi}e^{\frac{\varphi_1}{2}\Gamma^{34}}e^{-\log r w_r}\epsilon_3(x_-)\,,
\end{align}
Finally, for the remaining $x_-$ component of the Killing Spinor equation, we use that 
\begin{align}
W_{x_-}e^{-\psi W_\psi}e^{\frac{\varphi_1}{2}\Gamma^{34}}e^{-\log r w_r}=e^{-\psi W_\psi}e^{\frac{\varphi_1}{2}\Gamma^{34}}e^{-\log r w_r}w_{x_-}\,,
\end{align}
so that the full solution is given by
\begin{align}
     \epsilon=e^{-\psi W_\psi}e^{\frac{\varphi_1}{2}\Gamma^{34}}e^{-\log r w_r}e^{-x_-w_{x_-}}P_1P_2P_0\epsilon_0,
\end{align}
where $\epsilon_0$ is a constant spinor.

\bibliographystyle{utphys}
\bibliography{refs.bib}

@article{NEWPAPER,
    author = "Maurelli, S. and Noris, R. and Oyarzo, M. and Samtleben, H. and Saha, A and Trigiante, M.",
    title = "{Work in progress}",
}

@article{Bardeen:1973gs,
    author = "Bardeen, James M. and Carter, B. and Hawking, S. W.",
    title = "{The Four laws of black hole mechanics}",
    doi = "10.1007/BF01645742",
    journal = "Commun. Math. Phys.",
    volume = "31",
    pages = "161--170",
    year = "1973"
}

@book{Compere:2019qed,
    author = "Comp{\`e}re, Geoffrey",
    title = "{Advanced Lectures on General Relativity}",
    doi = "10.1007/978-3-030-04260-8",
    isbn = "978-3-030-04259-2, 978-3-030-04260-8",
    publisher = "Springer, Cham",
    address = "Cham, Switzerland",
    volume = "952",
    month = "2",
    year = "2019"
}

@book{Griffiths:2009dfa,
    author = "Griffiths, Jerry B. and Podolsky, Jiri",
    title = "{Exact Space-Times in Einstein's General Relativity}",
    doi = "10.1017/CBO9780511635397",
    isbn = "978-1-139-48116-8",
    publisher = "Cambridge University Press",
    address = "Cambridge",
    series = "Cambridge Monographs on Mathematical Physics",
    year = "2009"
}

@article{Castellani:2024ial,
    author = "Castellani, Federico and Nunez, Carlos",
    title = "{Holography for confined and deformed theories: TsT-generated solutions in type IIB supergravity}",
    eprint = "2410.00094",
    archivePrefix = "arXiv",
    primaryClass = "hep-th",
    doi = "10.1007/JHEP12(2024)155",
    journal = "JHEP",
    volume = "12",
    pages = "155",
    year = "2024"
}

@article{Catal-Ozer:2005dux,
    author = "Catal-Ozer, Aybike",
    title = "{Lunin-Maldacena deformations with three parameters}",
    eprint = "hep-th/0512290",
    archivePrefix = "arXiv",
    doi = "10.1088/1126-6708/2006/02/026",
    journal = "JHEP",
    volume = "02",
    pages = "026",
    year = "2006"
}

@article{Romans:1991nq,
    author = "Romans, L. J.",
    title = "{Supersymmetric, cold and lukewarm black holes in cosmological Einstein-Maxwell theory}",
    eprint = "hep-th/9203018",
    archivePrefix = "arXiv",
    reportNumber = "PRINT-92-0114 (JPL,CAL-TECH)",
    doi = "10.1016/0550-3213(92)90684-4",
    journal = "Nucl. Phys. B",
    volume = "383",
    pages = "395--415",
    year = "1992"
}

@book{Tomasiello:2022dwe,
    author = "Tomasiello, Alessandro",
    title = "{Geometry of String Theory Compactifications}",
    doi = "10.1017/9781108635745",
    isbn = "978-1-108-63574-5, 978-1-108-47373-6",
    publisher = "Cambridge University Press",
    year = "2022"
}

@article{Andrianopoli:2023dfm,
	archiveprefix = {arXiv},
	author = {Andrianopoli, L. and Cerchiai, B. L. and Noris, R. and Ravera, L. and Trigiante, M. and Zanelli, J.},
	doi = {10.1103/PhysRevD.108.044011},
	eprint = {2305.17168},
	journal = {Phys. Rev. D},
	number = {4},
	pages = {044011},
	primaryclass = {hep-th},
	title = {New torsional deformations of locally {AdS$_3$} space},
	volume = {108},
	year = {2023}}

@article{Detournay:2005fz,
	archiveprefix = {arXiv},
	author = {Detournay, Stephane and Orlando, Domenico and Petropoulos, P. Marios and Spindel, Philippe},
	doi = {10.1088/1126-6708/2005/07/072},
	eprint = {hep-th/0504231},
	journal = {JHEP},
	pages = {072},
	title = {Three-dimensional black holes from deformed anti-de {S}itter},
	volume = {07},
	year = {2005}}

@article{Israel:2004vv,
	archiveprefix = {arXiv},
	author = {Israel, Dan and Kounnas, Costas and Orlando, Domenico and Petropoulos, P. Marios},
	doi = {10.1002/prop.200410190},
	eprint = {hep-th/0405213},
	journal = {Fortsch. Phys.},
	pages = {73--104},
	reportnumber = {LPTENS-04-26, CPTH-RR014-0404},
	title = {Electric/magnetic deformations of {$S^3$} and {AdS$_3$}, and geometric cosets},
	volume = {53},
	year = {2005}}

@article{Detournay:2012pc,
	archiveprefix = {arXiv},
	author = {Detournay, Stephane and Hartman, Thomas and Hofman, Diego M.},
	doi = {10.1103/PhysRevD.86.124018},
	eprint = {1210.0539},
	journal = {Phys. Rev. D},
	pages = {124018},
	primaryclass = {hep-th},
	reportnumber = {SLAC-PUB-16086},
	title = {Warped conformal field theory},
	volume = {86},
	year = {2012}}

@article{Eberhardt:2019ywk,
	archiveprefix = {arXiv},
	author = {Eberhardt, Lorenz and Gaberdiel, Matthias R. and Gopakumar, Rajesh},
	doi = {10.1007/JHEP02(2020)136},
	eprint = {1911.00378},
	journal = {JHEP},
	pages = {136},
	primaryclass = {hep-th},
	title = {Deriving the {AdS$_{3}$/CFT$_{2}$} correspondence},
	volume = {02},
	year = {2020}}

@article{Strominger:1997eq,
	archiveprefix = {arXiv},
	author = {Strominger, Andrew},
	doi = {10.1088/1126-6708/1998/02/009},
	eprint = {hep-th/9712251},
	journal = {JHEP},
	pages = {009},
	reportnumber = {HUTP-97-A106},
	title = {{Black hole entropy from near horizon microstates}},
	volume = {02},
	year = {1998}}

@article{Lunin:2005jy,
	archiveprefix = {arXiv},
	author = {Lunin, Oleg and Maldacena, Juan Martin},
	doi = {10.1088/1126-6708/2005/05/033},
	eprint = {hep-th/0502086},
	journal = {JHEP},
	pages = {033},
	title = {Deforming field theories with {$U(1) \times U(1)$} global symmetry and their gravity duals},
	volume = {05},
	year = {2005}}

@article{Maldacena:1998bw,
	archiveprefix = {arXiv},
	author = {Maldacena, Juan Martin and Strominger, Andrew},
	doi = {10.1088/1126-6708/1998/12/005},
	eprint = {hep-th/9804085},
	journal = {JHEP},
	pages = {005},
	primaryclass = {hep-th},
	reportnumber = {HUTP-98-A016},
	slaccitation = {%%CITATION = HEP-TH/9804085;%%},
	title = {{A}d{S}$_3$ black holes and a stringy exclusion principle},
	volume = {12},
	year = {1998}}

@article{deBoer:1998kjm,
	archiveprefix = {arXiv},
	author = {de Boer, Jan},
	doi = {10.1016/S0550-3213(99)00160-1},
	eprint = {hep-th/9806104},
	journal = {Nucl. Phys.},
	pages = {139-166},
	primaryclass = {hep-th},
	reportnumber = {LBL-41931, LBNL-41931, UCB-PTH-98-32},
	slaccitation = {%%CITATION = HEP-TH/9806104;%%},
	title = {Six-dimensional supergravity on ${S}^3 \times {AdS}_3$ and $2d$ conformal field theory},
	volume = {B548},
	year = {1999}}

@article{Eberhardt:2017fsi,
	archiveprefix = {arXiv},
	author = {Eberhardt, Lorenz and Gaberdiel, Matthias R. and Gopakumar, Rajesh and Li, Wei},
	doi = {10.1007/JHEP03(2017)124},
	eprint = {1701.03552},
	journal = {JHEP},
	pages = {124},
	primaryclass = {hep-th},
	slaccitation = {%%CITATION = ARXIV:1701.03552;%%},
	title = {{BPS} spectrum on {AdS}$_3\times ${S}$^3 \times ${S}$^3 \times ${S}$^1$},
	volume = {03},
	year = {2017}}

@article{Giveon:1994fu,
	archiveprefix = {arXiv},
	author = {Giveon, Amit and Porrati, Massimo and Rabinovici, Eliezer},
	doi = {10.1016/0370-1573(94)90070-1},
	eprint = {hep-th/9401139},
	journal = {Phys. Rept.},
	pages = {77-202},
	primaryclass = {hep-th},
	reportnumber = {RI-1-94, NYU-TH-94-01-01},
	slaccitation = {%%CITATION = HEP-TH/9401139;%%},
	title = {Target space duality in string theory},
	volume = {244},
	year = {1994}}

@article{Fukuma:1999jt,
    author = "Fukuma, Masafumi and Oota, Takeshi and Tanaka, Hirokazu",
    title = "{Comments on T dualities of Ramond-Ramond potentials on tori}",
    eprint = "hep-th/9907132",
    archivePrefix = "arXiv",
    reportNumber = "YITP-99-43",
    doi = "10.1143/PTP.103.425",
    journal = "Prog. Theor. Phys.",
    volume = "103",
    pages = "425--446",
    year = "2000"
}

@article{Deger:1998nm,
	archiveprefix = {arXiv},
	author = {Deger, S. and Kaya, A. and Sezgin, E. and Sundell, P.},
	doi = {10.1016/S0550-3213(98)00555-0},
	eprint = {hep-th/9804166},
	journal = {Nucl.Phys.},
	pages = {110-140},
	primaryclass = {hep-th},
	reportnumber = {CTP-TAMU-15-98},
	slaccitation = {%%CITATION = HEP-TH/9804166;%%},
	title = {Spectrum of ${D} = 6$, ${N}=4b$ supergravity on {AdS}$_3\times {S}^3$},
	volume = {B536},
	year = {1998}}

@article{Strominger:1996sh,
	archiveprefix = {arXiv},
	author = {Strominger, Andrew and Vafa, Cumrun},
	doi = {10.1016/0370-2693(96)00345-0},
	eprint = {hep-th/9601029},
	journal = {Phys. Lett.},
	pages = {99-104},
	slaccitation = {%%CITATION = HEP-TH/9601029;%%},
	title = {{Microscopic Origin of the Bekenstein-Hawking Entropy}},
	volume = {B379},
	year = {1996}}

@article{Brown:1986nw,
	author = {Brown, J. David and Henneaux, M.},
	doi = {10.1007/BF01211590},
	journal = {Commun. Math. Phys.},
	pages = {207-226},
	slaccitation = {%%CITATION = CMPHA,104,207;%%},
	title = {{Central Charges in the Canonical Realization of Asymptotic Symmetries: An Example from Three-Dimensional Gravity}},
	volume = {104},
	year = {1986}}

@article{Hoare:2022asa,
    author = "Hoare, Ben and Seibold, Fiona K. and Tseytlin, Arkady A.",
    title = "{Integrable supersymmetric deformations of AdS$_{3}$\texttimes{} S$^{3}$\texttimes{} T$^{4}$}",
    eprint = "2206.12347",
    archivePrefix = "arXiv",
    primaryClass = "hep-th",
    reportNumber = "Imperial-TP-AT-2022-02",
    doi = "10.1007/JHEP09(2022)018",
    journal = "JHEP",
    volume = "09",
    pages = "018",
    year = "2022"
}

@article{Orlando:2010ay,
    author = "Orlando, Domenico and Uruchurtu, Linda I.",
    title = "{Warped anti-de Sitter spaces from brane intersections in type II string theory}",
    eprint = "1003.0712",
    archivePrefix = "arXiv",
    primaryClass = "hep-th",
    reportNumber = "IPMU10-0041, IMPERIAL-TP-2010-LIU-01",
    doi = "10.1007/JHEP06(2010)049",
    journal = "JHEP",
    volume = "06",
    pages = "049",
    year = "2010"
}

@article{Orlando:2019rjg,
    author = "Orlando, Domenico and Reffert, Susanne and Sekiguchi, Yuta and Yoshida, Kentaroh",
    title = "{O(d,d) transformations preserve classical integrability}",
    eprint = "1907.03759",
    archivePrefix = "arXiv",
    primaryClass = "hep-th",
    reportNumber = "KUNS-2767",
    doi = "10.1016/j.nuclphysb.2019.114880",
    journal = "Nucl. Phys. B",
    volume = "950",
    pages = "114880",
    year = "2020"
}

@article{Maurelli:2025iba,
    author = "Maurelli, S. and Noris, R. and Oyarzo, M. and Samtleben, H. and Trigiante, M.",
    title = "{Supersymmetric warped solutions from Type IIB orientifold reduction}",
    eprint = "2504.16822",
    archivePrefix = "arXiv",
    primaryClass = "hep-th",
    doi = "10.1007/JHEP08(2025)013",
    journal = "JHEP",
    volume = "08",
    pages = "013",
    year = "2025"
}

@article{Maldacena:2000hw,
    author = "Maldacena, Juan Martin and Ooguri, Hirosi",
    title = "{Strings in AdS(3) and SL(2,R) WZW model 1.: The Spectrum}",
    eprint = "hep-th/0001053",
    archivePrefix = "arXiv",
    reportNumber = "CALT-68-2245, CITUSC-99-010, HUTP-99-A027, LBNL-44375, UCB-PTH-99-48, LBL-44375",
    doi = "10.1063/1.1377273",
    journal = "J. Math. Phys.",
    volume = "42",
    pages = "2929--2960",
    year = "2001"
}

@article{Maldacena:2000kv,
    author = "Maldacena, Juan Martin and Ooguri, Hirosi and Son, John",
    title = "{Strings in AdS(3) and the SL(2,R) WZW model. Part 2. Euclidean black hole}",
    eprint = "hep-th/0005183",
    archivePrefix = "arXiv",
    reportNumber = "CALT-68-2266, CITUSC-00-021, HUTP-00-A009, UCB-PTH-00-10",
    doi = "10.1063/1.1377039",
    journal = "J. Math. Phys.",
    volume = "42",
    pages = "2961--2977",
    year = "2001"
}

@article{Maldacena:2001km,
    author = "Maldacena, Juan Martin and Ooguri, Hirosi",
    title = "{Strings in AdS(3) and the SL(2,R) WZW model. Part 3. Correlation functions}",
    eprint = "hep-th/0111180",
    archivePrefix = "arXiv",
    reportNumber = "CALT-68-2360, CITUSC-01-042",
    doi = "10.1103/PhysRevD.65.106006",
    journal = "Phys. Rev. D",
    volume = "65",
    pages = "106006",
    year = "2002"
}

@article{Georgescu:2025jlx,
    author = "Georgescu, Silvia",
    title = "{Non-linear asymptotic symmetries in warped AdS$_3$ holography}",
    eprint = "2507.00144",
    archivePrefix = "arXiv",
    primaryClass = "hep-th",
    month = "6",
    year = "2025"
}

@article{Georgescu:2024iam,
    author = "Georgescu, Silvia and Guica, Monica and Kovensky, Nicolas",
    title = "{Ascending the attractor flow in the D1-D5 system}",
    eprint = "2401.01298",
    archivePrefix = "arXiv",
    primaryClass = "hep-th",
    month = "1",
    year = "2024"
}

\end{document}